\documentclass[letterpaper]{article} 
\usepackage{aaai24}  
\usepackage{times}  
\usepackage{helvet}  
\usepackage{courier}  
\usepackage[hyphens]{url}  
\usepackage{graphicx} 
\usepackage{natbib}  
\usepackage{caption} 
\usepackage{algorithm}
\usepackage{algorithmic}
\usepackage{mathtools}
\usepackage{xcolor}
\usepackage{amssymb}
\usepackage[nice]{nicefrac}

\usepackage{newfloat}
\usepackage{listings}
\DeclareCaptionStyle{ruled}{labelfont=normalfont,labelsep=colon,strut=off} 
\floatstyle{ruled}
\newfloat{listing}{tb}{lst}{}
\floatname{listing}{Listing}
\title{Deep Quantum Error Correction}
\author{
    Yoni Choukroun,
    Lior Wolf
}
\affiliations{
    The Blavatnik School of Computer Science\\
    Tel Aviv University\\
    choukroun.yoni@gmail.com, wolf@cs.tau.ac.il
}

\usepackage{bibentry}

\begin{document}

\maketitle

\begin{abstract}
Quantum error correction codes (QECC) are a key component for realizing the potential of quantum computing. QECC, as its classical counterpart (ECC), enables the reduction of error rates, by distributing quantum logical information across redundant physical qubits, such that errors can be detected and corrected. In this work, we efficiently train novel {\emph{end-to-end}} deep quantum error decoders. We resolve the quantum measurement collapse by augmenting syndrome decoding to predict an initial estimate of the system noise, which is then refined iteratively through a deep neural network. The logical error rates calculated over finite fields are directly optimized via a differentiable objective, enabling efficient decoding under the constraints imposed by the code. Finally, our architecture is extended to support faulty syndrome measurement, by efficient decoding of repeated syndrome sampling. The proposed method demonstrates the power of neural decoders for QECC by achieving state-of-the-art accuracy, outperforming {for small distance topological codes,} the existing {end-to-end }neural and classical decoders, which are often computationally prohibitive.
Our code is added as supplementary material.
\end{abstract}

\section{Introduction}
Error Correcting Codes (ECC) are required in order to overcome computation and transmission corruption in almost every computation device \citep{shannon1948mathematical,mackay2003information}. 
Quantum systems are known for being extremely noisy, thereby requiring the use of error correction~\citep{lidar2013quantum,ballance2016high,huang2019fidelity,foxen2020demonstrating}.  However, adapting existing classical ECC methods to the quantum domain (QECC) is not straightforward \citep{raimond1996quantum}.

The first difficulty in applying ECC-based knowledge to QECC arises from the no-cloning theorem for quantum states \citep{wootters1982single}, which asserts that it is impossible to clone a quantum state and thus add arbitrarily redundant parity information, as done in classical ECC. 
The second challenge is the need to detect and correct quantum continuous bit-flips, as well as phase-flips  \citep{cory1998experimental,schindler2011experimental}, while classical ECC addresses only bit-flip errors. 
A third major challenge is the wave function collapse phenomenon: any direct measurements (while being standard in ECC) of the qubits would cause the wave function to collapse and erase the encoded quantum information \citep{neumann1955mathematical}. 
 
Shor \cite{shor1995scheme} proposed the first quantum error correction scheme, demonstrating that these challenges can be overcome. 
Subsequently, threshold \footnote{A threshold is the minimal error rate such that adding more physical qubits results in fewer logical errors.} theorems have shown that increasing the distance of a code will result in a corresponding reduction in the logical error rate, signifying that quantum error correction codes can arbitrarily suppress the logical error rate \citep{aharonov1997fault,kitaev1997quantum,preskill1998reliable}. 
This distance increase is obtained by developing encoding schemes that reliably store and process information in a logical set of qubits, by encoding it redundantly on top of a larger set of less reliable physical qubits \citep{nielsen2002quantum}. 
 
 Most current QECC methods fall in the category of stabilizer codes, which can be seen as a generalization of the classical linear codes \citep{gottesman1997stabilizer}. Similarly to the classical parity check constraints, a group of stabilizer operators can provide a syndrome, while preserving the logical quantum states and enabling error detection. 

Optimal decoding is defined by the unfeasible NP-hard maximum likelihood rule \citep{dennis2002topological,kuo2020hardnesses}. 
Considerable research dedicated to the design of codes with some additional algebraic structure have been done in order to support efficient decoding \citep{calderbank1996good,schindler2011experimental}. One of the most promising categories of codes are topological codes, particularly surface codes, which originated from Toric codes \citep{bravyi1998quantum,
kitaev1997quantum,kitaev1997quantum3,dennis2002topological,kitaev2003fault}. 
Given boundary conditions, the idea is to encode every logical qubit in a 2D lattice of physical qubits. 
This local design of the code via nearest-neighbors coupled qubits allows the correction of a wide range of errors,and, under certain assumptions, surface codes provide an exponential reduction in the error rate \citep{kitaev1997quantum,kitaev1997quantum2}. 

In this work, we present a novel {end-to-end} neural quantum decoding algorithm that can be applied to any stabilizer code.
{We first predict an approximation of the system noise, turning the quantum syndrome decoding task into a form that is compliant with the classical ECC Transformer decoder \cite{choukroun2022error}, which is based on the parity-check matrix. This initial estimate of the noise is reminiscent of Monte Carlo Markov Chains methods \cite{wootton2012high}, which start the decoding process with a random error that is compatible with the syndrome, and then refine it iteratively.} 
{Second, in order to support logical error optimization, we develop} a novel differentiable way of learning under the Boolean algebra constraints of the logical error rate metric. 
Finally, we propose a new architecture that is capable of performing quantum error correction under faulty syndrome measurements. As far as we can ascertain, this is the first time that (i) a Transformer architecture \citep{vaswani2017attention} is applied to the quantum syndrome decoding setting by augmenting syndrome decoding with noise prediction, (ii) a decoding algorithm is optimized directly over a highly non-differentiable finite field metric, (iii) a deep neural network is applied to the faulty syndrome decoding task in a time-invariant fashion.



 Applied to a variety of code lengths, noise, and {\color{black}measurement errors}, our method outperforms the state-of-the-art method. 
 This holds even when employing shallow architectures, providing a sizable speedup over the existing neural decoders, which remain  computationally inefficient \citep{edmonds1965paths,dennis2002topological,higgott2022pymatching}.


\section{Related Work}
Maximum Likelihood quantum decoding is an NP-hard problem \citep{kuo2020hardnesses} and several approximation methods have been developed as practical alternatives, trading off the accuracy for greater complexity. The Minimum-Weight Perfect Matching (MWPM) algorithm runs in polynomial time and is known to nearly achieve the optimal threshold for independent noise models \citep{bose1969combinatorial,micali1980v}. 
It is formulated as a problem of pairing excitation and can be solved using Blossom's algorithm \citep{kolmogorov2009blossom}. Approximations of this algorithm have been developed by parallelizing the group processing or removing edges that are unlikely to contribute to the matching process. 
However, MWPM implementations and modifications generally induce degradation in accuracy or remain too slow even for the current generation of superconducting qubits \citep{fowler2012towards,fowler2013minimum,meinerz2022scalable}.


Among other approaches, methods based on Monte Carlo Markov Chains (MCMC) \citep{wootton2012high,hutter2014efficient} iteratively modify the error estimate, to increase its likelihood with respect to the received syndrome. Renormalization Group methods \cite{duclos2010fast,duclos2013fault}  perform decoding by dividing the lattice into small cells of qubits. 
Union-Find decoders \citep{huang2020fault,delfosse2021almost} iteratively turn Pauli errors into losses that are corrected by the Union-Find data structure and are more suited to other types of noise. 

Multiple neural networks-based quantum decoders have emerged in the last few years  \citep{varsamopoulos2017decoding,torlai2017neural,krastanov2017deep,chamberland2018deep,andreasson2019quantum,wagner2020symmetries, sweke2020reinforcement,varona2020determination,meinerz2022scalable}. 
These methods are amendable to parallelization and can offer a high degree of adaptability. Current contributions make use of multi-layer perceptrons {\color{black}\citep{varsamopoulos2017decoding,torlai2017neural,krastanov2017deep,wagner2020symmetries}} or relatively shallow convolutional NNs {\color{black}\citep{andreasson2019quantum,sweke2020reinforcement}}, or couple \emph{local} neural decoding with classical methods for boosting the decoding accuracy {\color{black}\citep{meinerz2022scalable}}.

In parallel, deep learning methods have been improving steadily for classical ECC, reaching state-of-the-art results for {several code lengths.}
Many of these methods rely on augmenting the Belief-propagation algorithm with learnable parameters
\citep{pearl1988probabilistic,nachmani2016learning,lugosch2017neural,nachmani2019hyper,buchberger2020learned}, while others make use of more general neural network architectures \citep{cammerer2017scaling,gruber2017deep,kim2018communication,bennatan2018deep,choukroun2022denoising}.

Recently,  \cite{choukroun2022error} have proposed a transformer-based architecture \cite{vaswani2017attention} that is currently the state of the art in neural decoders for classical codes. 
We address QECC by expanding the ECCT architecture to account for the challenges arising from the transition from classical to quantum neural decoding. 

By using adapted masking obtained from the stabilizers, the Transformer based decoder is able to learn dependencies between related qubits.
However, it is \emph{important} to note that analogous expansions can be straightforward to apply to other neural decoder architectures.




\section{Background}
We provide the necessary background on classical and quantum error correction coding and a description of the state-of-the-art Error Correction Code Transformer (ECCT) decoder.
\paragraph{Classical Error Correction Code}
A linear code $C$ is defined by a binary generator matrix $G$ of size $k \times n$ and a binary parity check matrix $H$ of size $(n - k) \times n$ defined such that $GH^{T}=0$ over the order 2 Galois field $GF(2)$.

The input message $m \in \{0, 1\}^{k}$ is encoded by $G$ to a codeword $x \in C \subset \{0, 1\}^{n}$ satisfying $Hx=0$ and transmitted via a symmetric (potentially binary) channel,e.g., an additive white Gaussian noise (AWGN) channel. 
Let $y$ denote the channel output represented as $y=x_{s}+\varepsilon \in \mathcal{S} \subseteq \mathbb{R}^{n}$, where $x_s$ denotes the modulation of $x$, e.g. 
Binary Phase Shift Keying (BPSK), and $\varepsilon$ is random noise independent of the transmitted $x$.
The main goal of the decoder $f:\mathcal{S}\rightarrow \{0, 1\}^{n}$ is to provide an approximation of the codeword $\hat{x}=f(y)$. 

An important notion in ECC is the syndrome, which is obtained by multiplying the binary mapping of $y$ with the parity check matrix over $GF(2)$ such that
\begin{equation}
s\coloneqq s(y)=Hy_{b}\coloneqq H(x\oplus \varepsilon_{b})=H\varepsilon_{b},
\label{eq:syndrome_def}
\end{equation}
where $\oplus$ denotes the XOR operator, and $y_{b}$ and $\varepsilon_{b}$ denote the hard-decision vectors of $y$ and $\varepsilon$, respectively.

\textbf{Quantum Error Correction Code \quad}
The fundamental transition to the quantum realm is defined by the shift from the classical bit to the quantum bit (qubit), whose quantum state $|\psi\rangle$ is defined by 
\begin{equation}
|\psi\rangle = \alpha|0\rangle+\beta|1\rangle,
\ \ 
\text{s.t.} \ \ \alpha,\beta\in \mathbb{C}, \ |\alpha|^{2}+|\beta|^{2}=1
\label{eq:qubit}
\end{equation}
A coherent quantum error process $E$ can be decomposed into a sum of operators from the Pauli set $\{I, X, Z, XZ\}$, where the Pauli basis is defined by the identity mapping $I$, the quantum bit-flip $X$ and the phase-flip $Z$, such that
\begin{equation}
\begin{aligned}
I|\psi\rangle=&|\psi\rangle
\\
X|\psi\rangle=&\alpha X|0\rangle+\beta X|1\rangle= \alpha|1\rangle+\beta|0\rangle
\\
Z|\psi\rangle=&\alpha Z|0\rangle+\beta Z|1\rangle= \alpha|0\rangle-\beta|1\rangle
\end{aligned}
\label{eq:qerrors}
\end{equation}
and where the single-qubit error is defined as
\begin{equation}
\begin{aligned}
E|\psi\rangle=&\alpha_{I}|\psi\rangle+\alpha_{X}X|\psi\rangle+\alpha_{Z}Z|\psi\rangle+\alpha_{XZ}XZ|\psi\rangle
\end{aligned}
\label{eq:qerrors2}
\end{equation}
with $\alpha_{I},\alpha_{X},\alpha_{Z},\alpha_{XZ} \in \mathbb{C}$ being the expansion coefficients {\color{black}of the noise process}.

According to the no-cloning theorem, a quantum state $|\psi\rangle$ cannot be copied redundantly (i.e. $|\psi\rangle \otimes \dots \otimes |\psi\rangle$), where $\underbrace{\otimes \cdots \otimes}_\text{n}$ denotes the $n$-fold tensor product.
However, quantum information redundancy is possible through a \emph{logical} state encoding $|\psi\rangle_{n}$ of a given state $|\psi\rangle$ via quantum entanglement and a unitary operator $U$ such that $|\psi\rangle_{n}=U\big(|\psi\rangle \otimes |0\rangle \otimes \dots |0\rangle)\big)$. An example of such a unitary operator is the GHZ state \citep{greenberger1989going}, which is generated with CNOT gates.
In $|\psi\rangle_{n}$, the logical state is defined within a subspace of the expanded Hilbert space, which determines both the codespace $\mathcal{C}$ and its orthogonal error space $\mathcal{F}$ defined such that $E|\psi\rangle_{n} \in \mathcal{F}$.

The orthogonality {\color{black}of $\mathcal{C}$ and $\mathcal{F}$} makes it possible to determine the subspace occupied by the logical qubit through projective measurement, without compromising the encoded quantum information. 
In the context of quantum coding, the set $\mathcal{P}$ of non-destructive measurements of this type are called stabilizer measurements and are performed via additional qubits (ancilla bits). 
The result of all of the stabilizer measurements on a given state is called the \emph{syndrome}, such that for a given stabilizer generator $P\in \mathcal{P}$ we have $P|\psi\rangle_{n}=|\psi\rangle_{n}$, and $PE|\psi\rangle_{n}=-E|\psi\rangle_{n}$, $\forall |\psi\rangle_{n}\in \mathcal{C}$ given an anti-commuting (i.e. $-1$ eigenvalue) and thus detectable error $E$.
If the syndrome measurement is faulty, 
it might be necessary to repeat it to improve confidence in the outcome \citep[Section IV.B]{dennis2002topological}. 

An important class of Pauli operators is the class of logical operators.
These operators are not elements of the stabilizer group but commute with every stabilizer. While stabilizers operators act trivially in the code space, i.e. $P|\psi\rangle_{n}=|\psi\rangle_{n}$, logical operators {\color{black}$\ell \in \mathfrak{L}$} act non-trivially in it, i.e. $\exists |\phi\rangle_{n} \in \mathcal{C} \ \ s.t. \ \  \ell|\psi\rangle_{n}=|\phi\rangle_{n}$. 
Such operators commute with the stabilizers but can also represent undetectable errors \cite{QuantumErrorCorrection}.
Thus, similarly to the classical information bits, QECC benchmarks generally adopt logical error metrics, which measure the discrepancy between the predicted projected noise $\mathbb{L}\hat{\varepsilon}$ and the real one $\mathbb{L}{\varepsilon}$, {\color{black} where $\mathbb{L}$ is the discrete logical operators' matrix.}

\textbf{QECC from the ECC perspective\quad}
Another way to represent stabilizer codes is to split the stabilizer operators into two independent parity check matrices, defining the block parity check matrix $H$ 
such that
\mbox{$H=
\left(
    \begin{array}{c|c}
      H_{Z} & 0\\
      \hline
      0 & H_{X}
    \end{array}
    \right)
$}{\color{black}, separating phase-flip checks $H_Z$ and bit-flip checks $H_X$}. 
The syndrome $s$ is then computed as $s=H\varepsilon$, $H$ being the check-matrix defined according to the code stabilizers in $\mathcal{P}$, and $\varepsilon$ the binary noise.
The main goal of the quantum decoder {\color{black}$f:\{0, 1\}^{|\mathcal{P}|}\rightarrow \{0, 1\}^{|\mathfrak{L}|}$} is to provide a noise approximation given only the syndrome.

\begin{figure}[t]
\centering
  \includegraphics[trim={0 0 0 0},clip, width=1\linewidth]{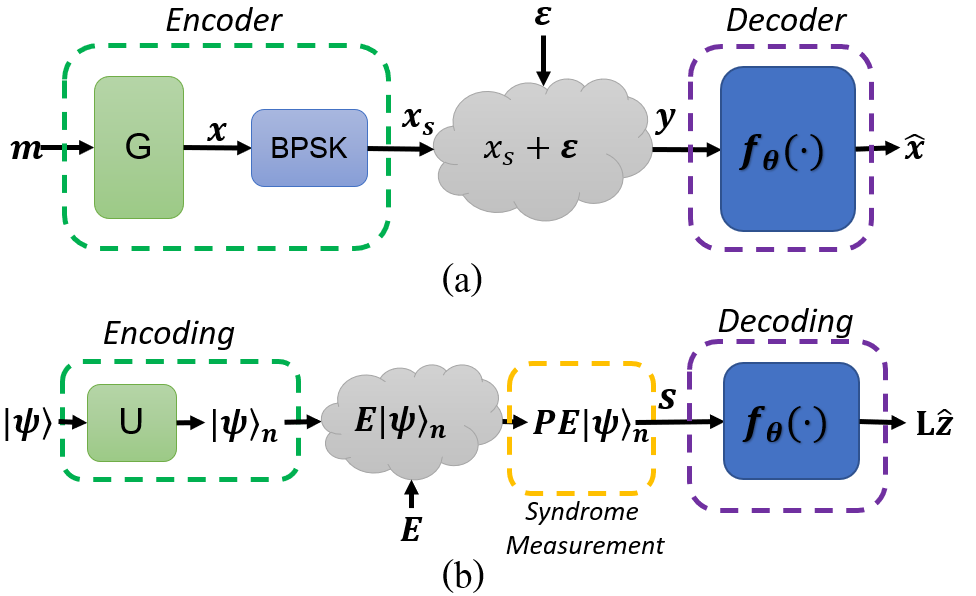}
  \caption{Illustration of the (a) classical and (b) quantum ECC system. Our work focuses on the latter. See Appendix A for a detailed illustration.}
  \label{fig:overview}

\end{figure}

Therefore, the quantum setting can be reduced to its classical counterpart as follows. The $k$ logical qubits are similar to the classical $k$ information bits, and the $n$ physical qubits are similar to the classical codeword. The syndrome of the quantum state can be computed or simulated similarly to the classical way, by defining the binary parity check matrix built upon the code quantum stabilizers. 
The main differences from classical ECC are: 
(i) no access to the current state is possible, while arbitrary measurement of $y$ is standard in the classical world;
(ii) we are interested in the logical qubits, predicting the code up to the logical operators mapping {\color{black}$\mathbb{L}$}; and finally,  
(iii) repetitive sampling of the syndrome due to the syndrome measurement error. {\color{black}These differences are at the core of our contributions.} 
{An illustration of the classical and quantum coding and decoding framework is given in Figure \ref{fig:overview}}. The goal of our method is to learn a decoder parameterized by weights $\theta$ such that
\mbox{$\hat{\varepsilon}=f_{\theta}(s)$}.

\textbf{Error Correction Code Transformer }
\ The SOTA ECCT \cite{choukroun2022error} has been recently proposed for classical error decoding.
It consists of a transformer architecture \cite{vaswani2017attention} with several modifications.

Following \cite{bennatan2018deep}, the model's input $h(y)$ is defined by the concatenation of the codeword-independent magnitude and syndrome, such that \mbox{$h(y)\coloneqq[|y|,1-2s(y)] \in \mathbb{R}^{2n-k}$} where $[\cdot,\cdot]$ denotes vector concatenation.
Each element is then embedded into a high-dimensional space for more expressivity, such that the initial positional embedding $\Phi$ is given by \mbox{$\Phi=\big( h(y)\cdot {1}_{d}^{T}\big) \odot W$}, where $W\in \mathbb{R}^{(2n-k)\times d}$ is the learnable embedding matrix and $\odot$ is the Hadamard product.

The interaction between the bits is performed naturally via the self-attention modules coupled with a binary mask derived from the parity-check matrix 
\begin{equation}
A_{H}(Q,K,V)=\text{Softmax}({d^{-\nicefrac{1}{2}}(QK^{T}+ g(H)}))V,
\end{equation}
where $g(H)$ is a binary masking function designed according to the parity-check matrix $H$, and $Q, K,V$ are the classical self-attention projection matrices.
 Masking enables the incorporation of sparse and efficient information about the code while avoiding the {\color{black} loop vulnerability of belief propagation-based decoders \cite{pearl1988probabilistic}.} Finally, the transformed embedding is projected onto a one-dimensional vector for noise prediction.
 The computational complexity is $O(N(d^2 n+ hd))$, where $N$ denotes the number of layers, and $n$ is the code length. 
 Here, $h << n^2$ denotes the number of elements in the mask, generally very small for sparse codes, including Toric and Surface codes.


\section{Method}
We present in this Section the elements of the proposed decoding framework, the complete architecture, and the training procedure.
{\color{black} 
 From now on, the binary {\color{black}block} parity check matrix {\color{black} is denoted by $H$, the binary noise by $\varepsilon$, the syndrome by $s=H\varepsilon$, and the logical operators' binary matrix by $\mathbb{L}$.}
 }
 

\paragraph{Overcoming Measurement Collapse by Prediction}
While syndrome decoding is a well-known procedure in ECC, most popular decoders, and especially neural decoders, assume the availability of arbitrary measurements of the channel's output. 
In the QECC setting, only the syndrome is available, since classical measurements are not allowed due to the wave function collapse phenomenon.

We thus propose to extend ECCT, by replacing the magnitude {\color{black}of the channel output $y$} with an initial estimate of the noise to be further refined by the code-aware network. In other words, we replace the channel's output magnitude measurement $h(y)=[|y|,1-2s]$ by ${h}_{q}(s)=[g_{\omega}(s),s]$. 

Denoting the ECCT decoder by $f_{\theta}$ we have
\begin{equation}
\hat{z}=f_{\theta}\big({h}_{q}(s)\big)=f_{\theta}\big([g_{\omega}(s),s]\big),
\end{equation}
where $g_{\omega}:\{0,1\}^{n_{s}}\rightarrow \mathbb{R}^{n}$ is the initial noise estimator, given by a shallow neural network parameterized by $\omega$. 
This way, {\color{black}the ECCT can process the estimated input and perform decoding by analyzing the input-syndrome interactions via the masking}. 
As a non-linear transformation of the syndrome, $g_{\omega}(s)$ is also independent of the quantum state/codeword and thus robust to overfitting. The estimator $g_{\omega}(s)$ is trained via the following objective
\begin{equation}
\mathcal{L}_{g} = \text{BCE}\big(g_{\omega}(s),\varepsilon\big),
\end{equation}
where $\text{BCE}$ is the binary cross entropy loss.
{\color{black}This shift from magnitude to initial error estimation is crucial for overcoming quantum measurement collapse and, as explored in the analysis section, leads to markedly better performance.}


\paragraph{Logical Decoding}
\label{sec:logical}





Contrary to classical ECC, quantum error correction aims to restore the noise up to a \emph{logical} generator of the code, such that several solutions can be valid error correction.
 Accordingly, the commonly used metric is the logical error rate (LER), which provides valuable information on the practical decoding performance {\color{black}\cite{lidar2013quantum}}.
Given the code's logical operator in its matrix form $\mathbb{L} \in \{0,1\}^{k \times n}$, we wish to minimize the following LER objective 
{\color{black}
\begin{equation}
\mathcal{L}_{\text{LER}} = \text{BCE}\big(\mathbb{L}f_{\theta}(s),\mathbb{L}\varepsilon\big).
\end{equation}
}
where the multiplications are performed over $GF(2)$ (i.e. binary modulo 2) and are thus highly non-differentiable.

We propose to optimize the objective using a differentiable equivalence mapping of the XOR (i.e., sum over $GF(2)$) operation as follows.
Defining the bipolar mapping \mbox{\color{black}$\phi:\{0,1\}\rightarrow \{\pm 1\}$} over $GF(2)$ as \mbox{$\phi(u)=1-2u, u\in \{0,1\}$}, {\color{black} we obtain the following  property} \mbox{$\phi(u\oplus v)=\phi(u)\phi(v), \forall u,v\in \{0,1\}$}.
Thus, with $\mathbb{L}_{i}$ the $i$-th row of $\mathbb{L}$ and $x$ a binary vector, we have {\color{black}$\forall i \in \{1\dots k\}$
\begin{equation}
\label{eq:polar-eq}
\big(\Lambda(\mathbb{L},x)\big)_{i} \coloneqq \mathbb{L}_{i}\oplus x=
\phi^{-1}\bigg(\Pi_{j} \phi\big((\mathbb{L})_{ij}\cdot x_{j}\big)\bigg).
\end{equation}
Thus, as a composition of differentiable functions $\Lambda(\mathbb{L},x)$ is differentiable and we can redefine our objective as folllows
\begin{equation}
\label{eq:logical-bin}
\mathcal{L}_{\text{LER}} = \text{BCE}\bigg(\Lambda\big(\mathbb{L},{bin}(f_{\theta}(s))\big),\mathbb{L}\varepsilon\bigg),
\end{equation}
}
where ${bin}$ denotes the binarization of the soft prediction of the trained model.
While many existing works make use of the straight-through estimator (STE) \citep{bengio2013estimating} for the binary quantization of the activations, we opt for its differentiable approximation with the sigmoid function (i.e., ${bin}(x)=\sigma(x)=(1+e^{-x})^{-1}$).
{\color{black}As shown in our ablation analysis, the performance of the STE is slightly inferior to the sigmoid approach.}

In addition to directly minimizing the LER metric, we are interested in noise prediction solutions that are close to the real system noise.
We, therefore, suggest regularizing the objective with the classical and popular Bit Error Rate (BER) objective defined as
$\mathcal{L}_{\text{BER}} = \text{BCE}\big(f_{\theta}(s),\varepsilon\big)$.
Combining the loss terms, the overall objective is given by  
\begin{equation}
\mathcal{L} = \lambda_{\text{BER}}\mathcal{L}_{\text{BER}}+\lambda_{\text{LER}}\mathcal{L}_{\text{LER}}+\lambda_{g}\mathcal{L}_{g}
\label{eq:qecctobjective}, 
\end{equation}
where $\lambda_{\text{BER},\text{LER},g}$ denote the weights of each objective. 

\paragraph{Noisy Syndrome Measurements}
In the presence of measurement errors, each syndrome measurement is repeated $T$ times.
{\color{black}This gives the decoder input an additional time dimension.}
Formally, given binary system noises $\{\varepsilon_{t}\}_{t=1}^{T}$ and binary measurement noises  $\{\tilde{\varepsilon}_{t}\}_{t=1}^{T}$, we have the syndrome $s_{t}$ at a given time $t\in \mathbb{N}_{+}$ defined as
\begin{equation}
s_{t} = \big(H(x\oplus \varepsilon_{1}\oplus \dots \oplus \varepsilon_{t}   ) \big)\oplus \tilde{\varepsilon}_{t}
\label{eq:noisy-synd}
\end{equation}

To remain invariant to the number of measurements, we first analyze each measurement separately and then perform global decoding by applying a symmetric pooling function, e.g. an average, in the middle of the neural decoder.

Given a NN decoder with $N$ layers and 
the hidden activation tensor $\varphi \in \mathbb{R}^{T\times n\times d_{l}}$ at layer $l=\lfloor N/2 \rfloor$, the new 
pooled embedding is given by summation along the first dimension $\tilde{\varphi} = \frac{1}{T}\sum_{t=1}^{T}\varphi_{t}$. 
The loss $\mathcal{L}_{g}$ is thus defined as the distance between the pooled embedding and the noise, i.e., 
\begin{equation}
\mathcal{L}_{g} = \text{BCE}\big({\sum_{t}g_{\omega}(s_{t})}/{T},\varepsilon\big) 
\end{equation}
with $\varepsilon$ the cumulative binary system noise.

{\color{black}Since we extend the ECCT architecture to its quantum counterpart (QECCT), the hidden activation tensor is now of shape $\varphi \in \mathbb{R}^{T\times h\times n\times d_{l}}$ with $h$ the number of self-attention heads, and pooling is performed at the $\lfloor N/2 \rfloor$ Transformer block.
}
{ An advantage of this approach is its low computational cost since the analysis and comparison are performed in parallel at the embedding level.}


\paragraph{Architecture and Training}
The initial encoding is defined as a $d$ dimensional one-hot encoding of the $n+n_{s}$ input elements where $n$ is the number of physical qubits and $n_{s}$ the length of the syndrome.
The network $g_{\omega}$ is defined as a shallow network with two fully connected layers of hidden dimensions equal to $5n_{s}$ and with a GELU non-linearity. The decoder is defined as a concatenation of $N$ decoding layers composed of self-attention and feed-forward layers interleaved with normalization layers. In case of faulty syndrome measurements, the $\lfloor N/2 \rfloor$th layer performs average pooling over the time dimension.

\begin{figure}[t]
\centering
    \includegraphics[trim={0 0 0 0},clip, width=1\linewidth]{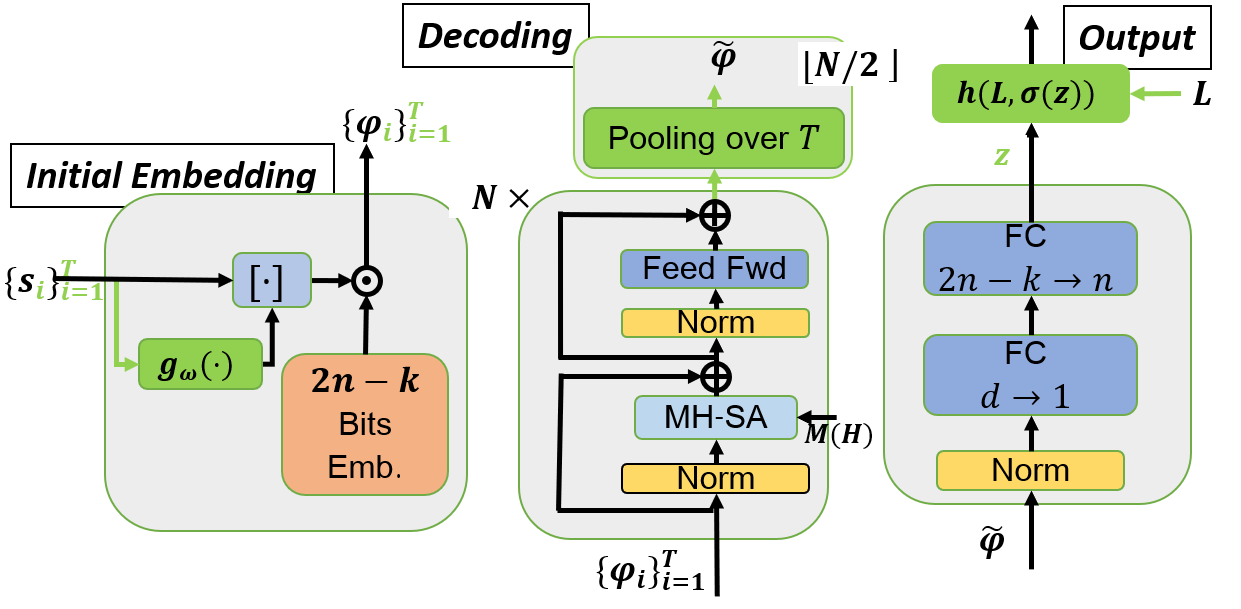}
  \caption{The proposed QECCT. The pooling layer is performed  after the middle self-attention block of the model. $M(H)$ is the mask derived from the parity-check matrix.}
\label{fig:qecct}
\end{figure}

The output {\color{black}is obtained via} two fully connected layers.
The first layer reduces the element-wise embedding to a one-dimensional $n+n_{s}$ vector and the second to an $n$ dimensional vector representing the soft decoded noise, trained over the objective given Eq. \ref{eq:qecctobjective}.
An illustration of the proposed QECCT is given in Figure \ref{fig:qecct}.

The complexity of the network is linear with the code length $n$ and quadratic with the embedding dimension $d$ and is defined by $\mathcal{O}(Nd^{2}n)$ for sparse (e.g. topological) codes. 
The acceleration of the proposed method (e.g. pruning, quantization, distillation, low-rank approximation) \cite{wang2020linformer,lin2021survey} is out of the scope of this paper and left for future work. 
For example, low-rank approximation would make the complexity also linear in $d$ \cite{wang2020linformer}. Also, no optimization of the sparse self-attention mechanism was employed in our implementation.





The Adam optimizer \cite{kingma2014adam} is used with 512 samples per minibatch, for 200 to 800 epochs depending on the code length, with 5000 minibatches per epoch.
The {training is performed by randomly sampling noise in the physical error rate testing range.}
The default weight parameters are  \mbox{$\lambda_{g}=0.5,\lambda_{\text{LER}}=1,\lambda_{\text{BER}}=0.5$}. 
Other configurations and longer training can be beneficial but were not tested due to a lack of computational resources.
{\color{black}
The default architecture is $N=6,d=128$.
}

We initialized the learning rate to $5\cdot10^{-4}$ coupled with a cosine decay scheduler down to $5\cdot 10^{-7}$ at the end of training. 
No warmup \cite{xiong2020layer} was employed.

Training and experiments were performed on a 12GB Titan V GPU.
{\color{black}The training time ranges from 153 to 692 seconds per epoch for the 32 to 400 code lengths, respectively with the default architecture.
Test time is in the range of 0.1 to 0.6 milliseconds per sample.
The number of testing samples is set to $10^{6}$, enough to obtain a small standard deviation ($\sim10^{-4}$) between experiments}.

\subsection{Application to Topological Codes}
While our framework is universal in terms of the code, we focus on the popular \emph{Surface} codes and, {\color{black} more specifically, on Toric codes \cite{kitaev1997quantum,kitaev1997quantum3,kitaev2003fault}, which are their variant with periodic boundary conditions}.
These codes are among the most attractive candidates for quantum computing experimental realization, as they can be implemented on a two-dimensional grid of qubits with local check operators \cite{bravyi2018correcting}.
The physical qubits are placed on the edges of a two-dimensional lattice of length $L$, such that the stabilizers are defined with respect to the code lattice architecture that defines the codespace, where $k=2, n=2L^{2}$. 

The stabilizers are defined in two groups: vertex operators are defined on each \emph{vertex} as the product of $X$ operators on
the adjacent qubits and \emph{plaquette} operators are defined on each face as the product of $Z$ operators on the bordering qubits. 
Therefore, there exist a total of $2L^{2}$ stabilizers, $L^{2}$ for each stabilizer group. Assuming that a qubit is associated with every edge of the lattice, for a given vertex $v$ we have the vertex operator defined as $X_{v}=\Pi_{i\in v}X_{i}$, and for a given plaquette $p$, the plaquette operator defined as $Z_{p}=\Pi_{i \in p}Z_{i}$. An illustration is given in Appendix G. 

\begin{figure}[t]
\centering
\noindent  \begin{tabular}{@{}c@{}c@{}c@{}c@{}}
  \includegraphics[trim={0.4cm 1.8cm 0.3cm 1.8cm},clip, height=0.168249\linewidth]{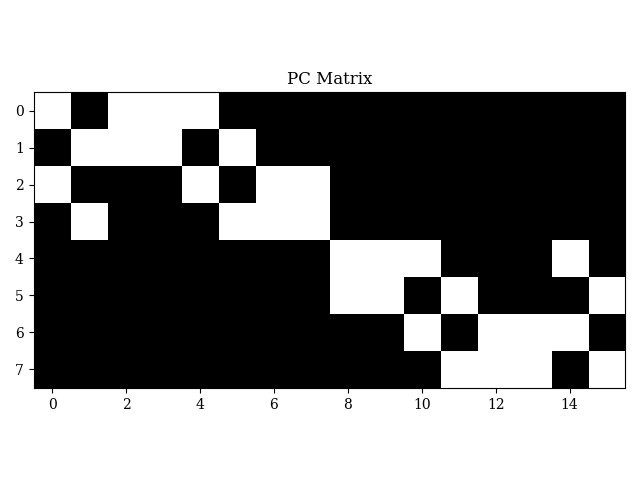}&
    \includegraphics[trim={0.4cm 0.4cm 0.6cm 0.4cm},clip, height=0.168249\linewidth]{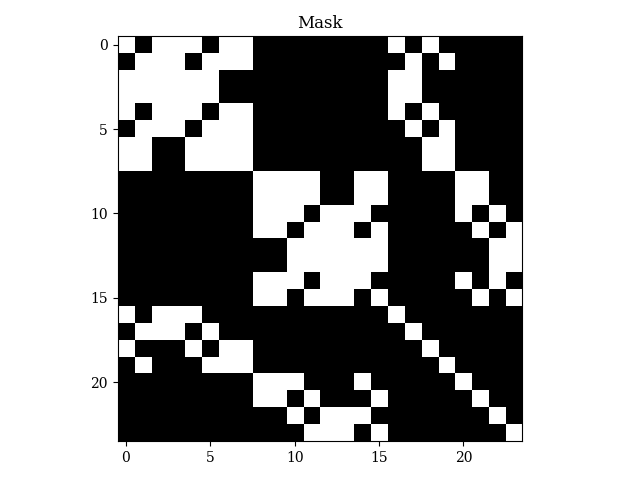} &
      \includegraphics[trim={0.5cm 1.8cm 0.3cm 1.8cm},clip, height=0.168249\linewidth]{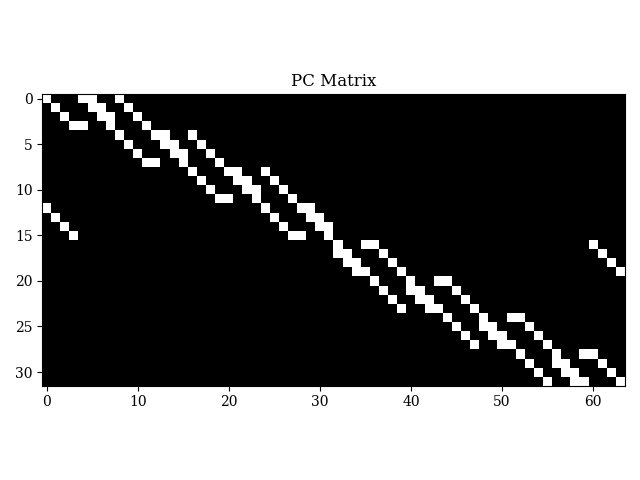}&
    \includegraphics[trim={0.4cm 0.4cm 0.0cm 0.4cm},clip, height=0.168249\linewidth]{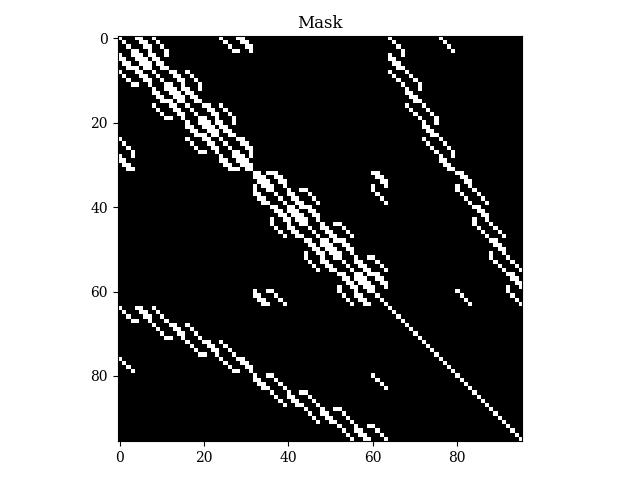}\\
    (a) & (b) & (c) & (d)\\
      \end{tabular}
  \caption{(a) The parity-check matrix and (b) the
induced masking for the 2-Toric code. (c,d) the corresponding matrix and mask for the 4-Toric code.
The parity-check matrix comprises two block matrices for the $X$ and $Z$ stabilizers.
We can observe the high sparsity of the Toric code. 
}
\label{fig:illus-pc-mask}
\end{figure}

The mask is defined such that the self-attention mechanism only takes into consideration bits related to each other in terms of the stabilizers (i.e., the parity-check matrix).
The parity-check matrices and their corresponding masks for several Toric codes are provided in Figure \ref{fig:illus-pc-mask}, where one can observe the high locality induced by the code architecture (the mask only reflects stabilizers-related elements).

\section{Experiments}
We evaluate our method on various Toric code {lengths}, considering the two common noise models: independent and depolarization. 
In independent (uncorrelated) noise, $X$ and $Z$ errors occur independently and with equal probabilities; therefore, decoding can be performed on the $X$ or $Z$ stabilizers separately. Depolarization noise assigns
equal probability $p/3$ to all three Pauli operators \mbox{$\mathbb{P}(X)=\mathbb{P}(Z)=\mathbb{P}(Y)=p/3, \mathbb{P}(I)=1-p$}, where $Y$ is the Pauli operator defined as $Y=iXZ$.

In the experiments where measurement errors are incorporated, each syndrome measurement is repeated $T=n$ times and the probability of the measurement error is the same as the probability of the syndrome error, i.e., the distributions of $\varepsilon$ and $\tilde{\varepsilon}$ in Eq. \ref{eq:noisy-synd} are the same as in \cite{dennis2002topological,higgott2022pymatching,wang2003confinement}.


The implementation of the Toric codes is taken from  \cite{krastanov2017deep}. As a baseline, we consider the Minimum-Weight Perfect Matching (MWPM) algorithm with the complexity of $\mathcal{O}((n^{3}+n^{2})log(n))$, also known as the Edmond or Blossom algorithm, which is the most popular decoder for topological codes. Its implementation is taken from \citep{pymatchingv2,higgott2022pymatching}  {and is close to quadratic average complexity.}
In the experiments, we employ code lengths similar to those for which the existing {end-to-end} neural decoders were tested, i.e., $2< L\leq 10$~\citep{varsamopoulos2017decoding,torlai2017neural,krastanov2017deep,chamberland2018deep,andreasson2019quantum,wagner2020symmetries}. It is worth noting that none of these previous methods outperform MWPM. The physical error ranges are taken around the thresholds of the different settings, as reported in  \cite{dennis2002topological,wang2003confinement,krastanov2017deep}.
On the tested codes and settings, the Union-find decoder \cite{park2022scalablecode} was not better than the MWPM algorithm.

As metrics, we present both the bit error rates (BER) and the logical error rate (LER), see {the Logical Decoding} section. The LER metric here is a word-level error metric, meaning there is an error if at least one qubit is different from the ground truth.



\subsection{Results}
\begin{figure}[t]
\centering
\noindent  \begin{tabular}{@{}ccc@{}}
  \includegraphics[trim={0 0 0 0},clip, width=0.49\linewidth]{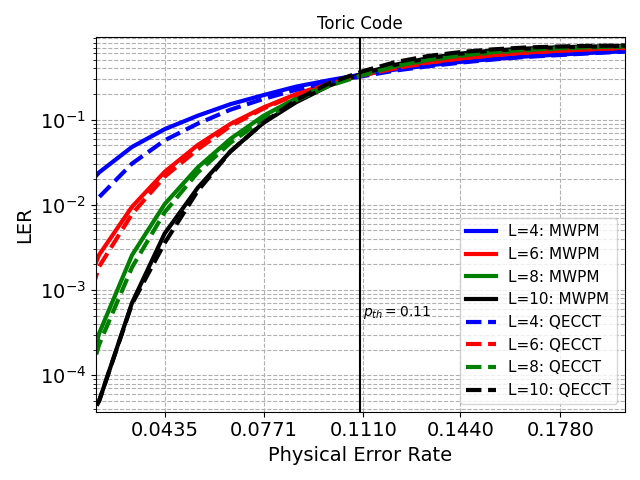}&
    \includegraphics[trim={0 0 0 0},clip, width=0.49\linewidth]{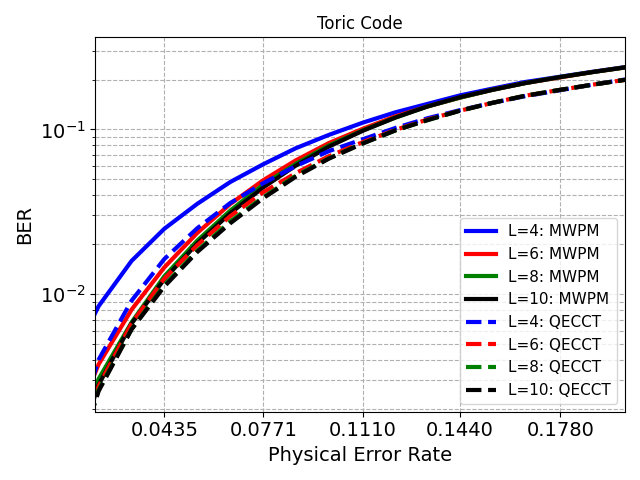}&
      \end{tabular}
  \caption{LER and BER performance for various physical error rates and lattice length on the Toric code with \emph{independent} noise and \emph{without} faulty syndrome measurements.}
\label{fig:Toric-norep-indepedent}
\end{figure}

\begin{figure}[t]
\centering
\noindent  \begin{tabular}{@{}ccc@{}}
  \includegraphics[trim={0 0 0 0},clip, width=0.49\linewidth]{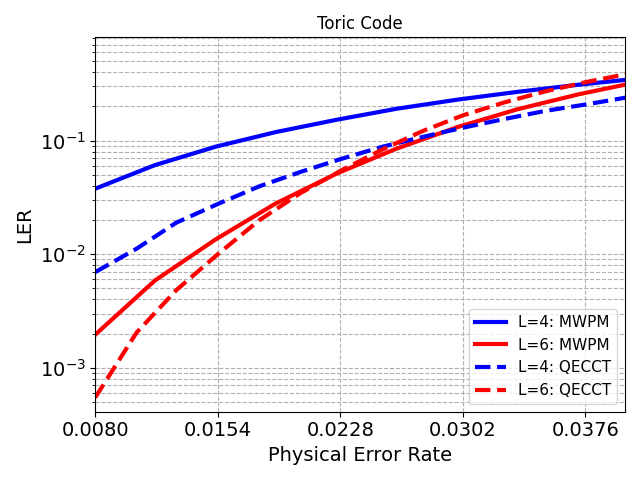}&
    \includegraphics[trim={0 0 0 0},clip, width=0.49\linewidth]{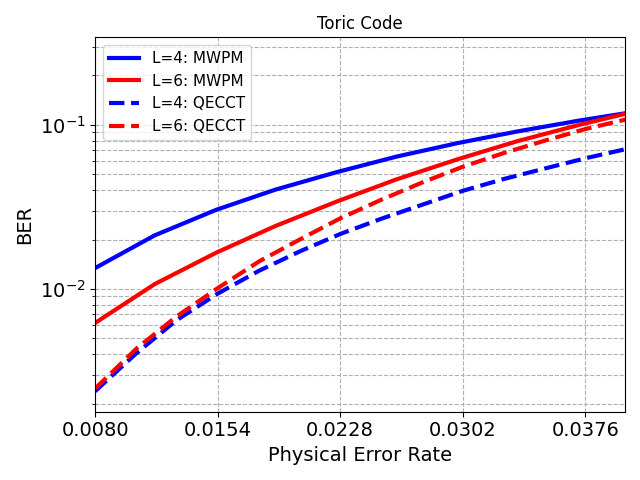}&
      \end{tabular}
  \caption{LER and BER performance for various physical error rates and lattice length on the Toric code with \emph{independent} noise and \emph{with} faulty syndrome measurements.}
\label{fig:Toric-withrep-indepedent}
\end{figure}

\begin{figure}[t]
\centering
\noindent  \begin{tabular}{@{}ccc@{}}
  \includegraphics[trim={0 0 0 0},clip, width=0.49\linewidth]{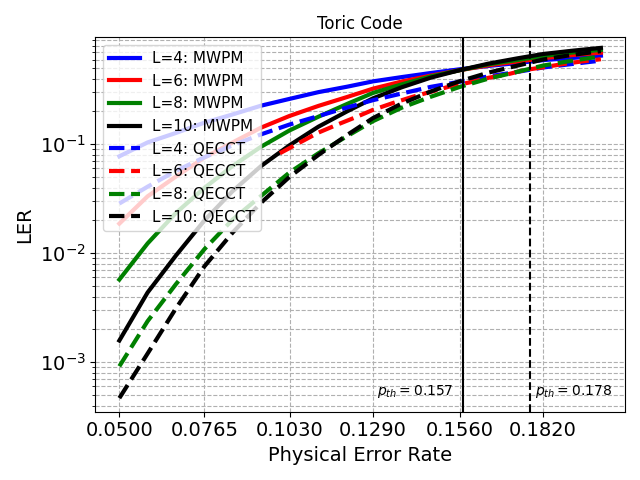}&
    \includegraphics[trim={0 0 0 0},clip, width=0.49\linewidth]{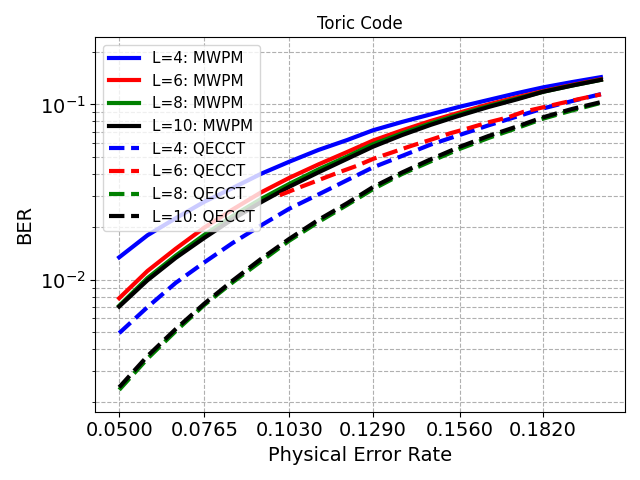}&
      \end{tabular}
  \caption{LER and BER performance for various error physical rates and lattice length on the Toric code with \emph{depolarization} noise and \emph{without} faulty syndrome measurements.}
\label{fig:Toric-norep-depol}
\end{figure}

\begin{figure}[t]
\centering
\noindent  \begin{tabular}{@{}ccc@{}}
  \includegraphics[trim={0 0 0 0},clip, width=0.49\linewidth]{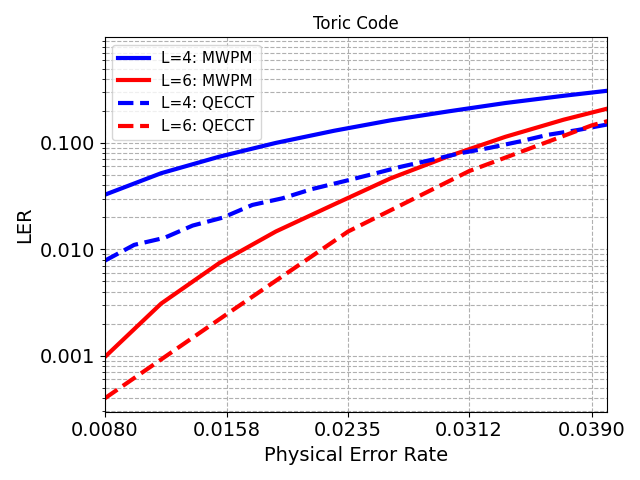}&
    \includegraphics[trim={0 0 0 0},clip, width=0.49\linewidth]{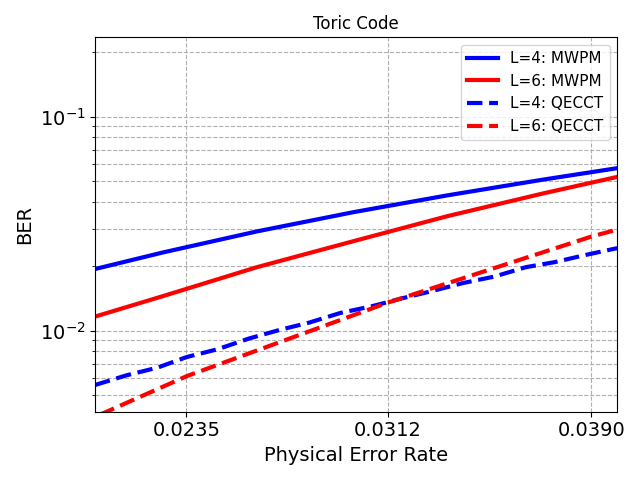}&
      \end{tabular}
  \caption{LER and BER performance for various error physical rates and lattice length on the Toric code with \emph{depolarization} noise and \emph{with} faulty syndrome measurements.}
\label{fig:Toric-withrep-depol}
\end{figure}

Figure \ref{fig:Toric-norep-indepedent} depicts the performance of the proposed method and the MWPM algorithm for different Toric code lengths under the \emph{independent} noise model and \emph{without} noisy measurements. Figure \ref{fig:Toric-withrep-indepedent} presents a similar comparison \emph{with} noisy measurements {\color{black}with $T=L$ and uniformly distributed syndrome error}.
Figure \ref{fig:Toric-norep-depol} and \ref{fig:Toric-withrep-depol} compare our method with MWPM for the \emph{depolarized} noise model, with and without noisy measurements, respectively. {We also provide the obtained threshold values.}

As can be seen, the proposed QECCT outperforms the SOTA MWPM algorithm by a large margin: (i) QECCT outperforms the MWPM algorithm on independent noise, where MWPM is known to almost reach  ML's threshold \cite{dennis2002topological}, and (ii) QECCT outperforms the SOTA MWPM on the challenging depolarization noise setting  by a large margin, where the obtained threshold is 0.178 compared to 0.157 for MWPM  and 0.189 for ML \cite{bombin2012strong}.
The very large gaps in BER imply that the proposed method is able to better detect exact corruptions.
The threshold is slightly lower for $L=10$ with depolarization noise while the BER is much lower, denoting a potential need for tuning the regularization parameter for larger code. 
In Appendix B we present performance statistics for the vanilla MLP decoder with similar capacity, highlighting both the contribution of our architecture and the importance of the proposed regularization term.

\begin{figure}[h]
\centering
\noindent  \begin{tabular}{@{}ccc@{}}
  \includegraphics[trim={0 0 0 0},clip, width=0.49\linewidth]{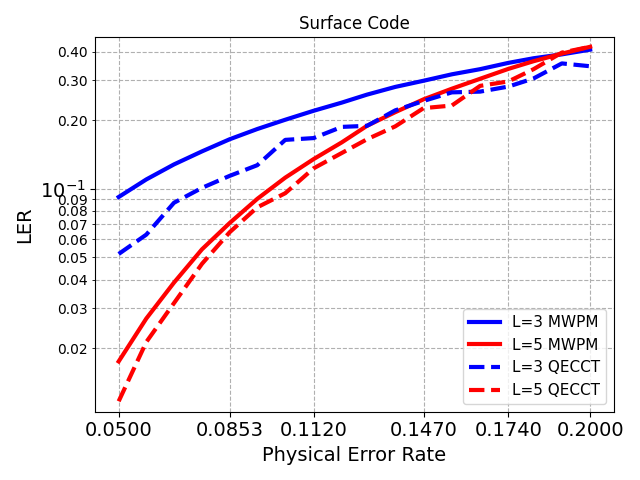}&
    \includegraphics[trim={0 0 0 0},clip, width=0.49\linewidth]{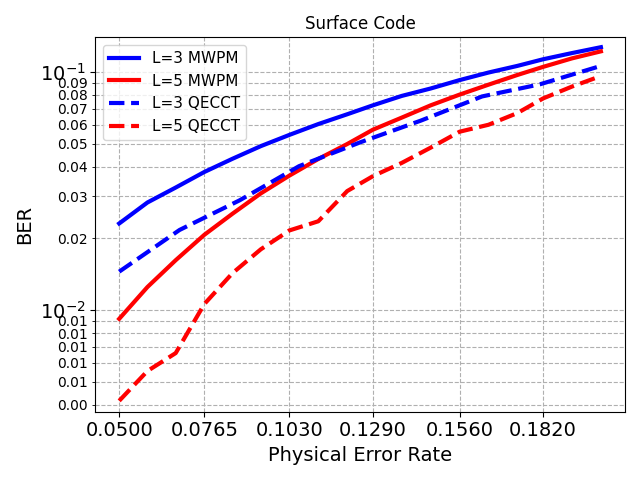}&
      \end{tabular}
  \caption{LER and BER performance for various error physical rates and lattice length on the Surface code with \emph{depolarization} noise. 
  }
\label{fig:surface-norep-depol}
\end{figure}

\begin{figure}[h]
\centering
\noindent  \begin{tabular}{@{}cc@{}}
  \includegraphics[trim={0 0 0 0},clip, width=0.49\linewidth]{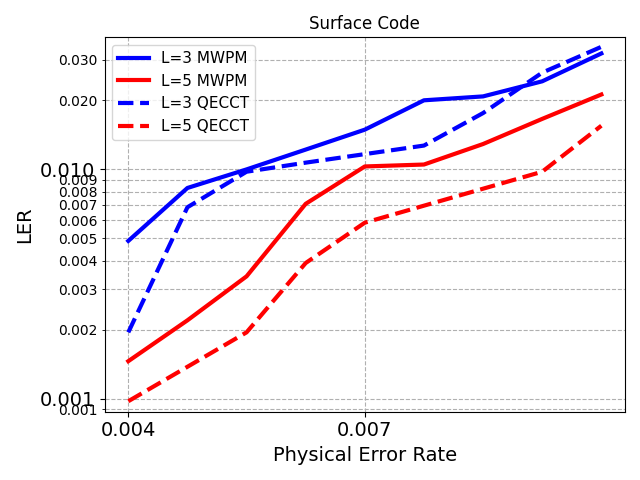}&
    \includegraphics[trim={0 0 0 0},clip, width=0.49\linewidth]{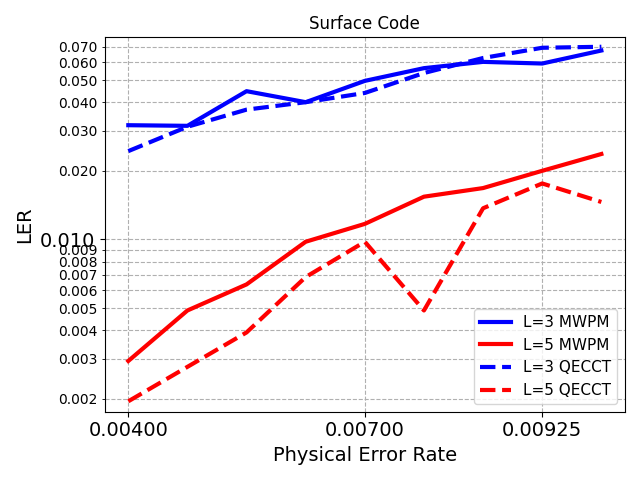}\\
    (a) & (b) \\
      \end{tabular}
  \caption{LER performance for various error physical rates and lattice length on the Surface code with \emph{circuit} noise \emph{without} repetitions.
  (a) Performance on randomly simulated syndromes. (b) Performance on randomly simulated syndromes with detectable noise only 
  (i.e. $s!=0$). 
  }
\label{fig:surface-norep-circuit}
\end{figure}


\subsection{Further Validation and an Ablation Study}

We extend our experiments beyond what is commonly done in the relevant ML work.  To check that the method is applicable not just for Toric codes, Figure \ref{fig:surface-norep-depol} shows the {\bf performance for different \emph{Surface} code} lengths of the proposed method and the MWPM algorithm under the depolarization noise model. The same parameters as were used for Toric codes are used here. As can be seen, the method is able to {similarly outperform MWPM for other codes as well}. The large gap in BER in favor of the QECCT probably means that the gap in LER can be made larger with other hyperparameters of the objective.

To explore the generality with respect to noise models, Figure \ref{fig:surface-norep-circuit} shows the performance {\bf under the \emph{circuit} noise model} of the proposed method and the MWPM algorithm for different \emph{Surface} code lengths. The channel is simulated using the STIM \citep{gidney2021stim} simulator of quantum stabilizer circuits where the same depolarization error probability is applied after every single and two-qubit Clifford operation, before every stabilizer measurement, and before the syndrome measurement.
Evidently, the proposed method is able to consistently outperform MWPM for this type of channel noise as well.

\noindent{\bf The impact of the noise estimator $g_{\omega}$} is studied in Appendix C, where it is shown that the initial noise estimator is critical for performance. 
{\bf The impact of the various objectives\quad} in Eq. \ref{eq:qecctobjective} is provided in Appendix D. 
It is demonstrated via the gradient norm that training solely with the $\mathcal{L}_{LER}$ objective converges rapidly to a bad local minimum.
Training with $\mathcal{L}_{BER}$ only produces far worse results than combining it with the $\mathcal{L}_{LER}$ objective. For high SNR, a model trained with $\mathcal{L}_{BER}$ objective yields a {\color{black}27 times higher LER} than the model trained with combined objectives. Moreover, optimizing with the noise estimator objective $\mathcal{L}_{g}$ results, for high SNR, in a 46\% improvement over not employing regularization.  
\noindent{\bf The impact of Pooling}
is explored in Appendix E, where various scenarios are compared. 
Empirical evidence is provided to support pooling in the middle layer, as suggested. 
Finally, \noindent{\bf the impact of the mask and the architecture} is explored in Appendix F. Specifically, masking, the model's capacity, and the STE as ${bin}$ function from Eq \ref{eq:logical-bin} are being evaluated. We can observe that while less important than with classical codes \citep{choukroun2022error}, the mask still substantially impacts performance. Also, we note increasing the capacity of the network enables better representation and decoding. 

\noindent{\bf Limitations}
While the proposed method has a smaller complexity than classical SOTA, its implementation in a straightforward way makes it difficult to train for larger codes or more repetitions, e.g. unstructured sparse self-attention is not easy to implement on general-purpose DL accelerators.   
{Larger architectures and longer training times would enable larger code correction and are expected to also improve accuracy, deepening the gap from other methods.  
As a point of reference, GPT-3 \citep{brown2020language} successfully operates on 2K inputs with a similar Transformer model but with $N=96,d=12K$. }

\section{Conclusion}
We present a novel Transformer-based framework for decoding quantum codes, offering multiple technical contributions enabling the effective representation and training under the QECC constraints.
First, the framework helps overcome the measurement collapse phenomenon by predicting the noise and then refining it. 
Second, we present a novel paradigm for differentiable training of highly non-differentiable functions, with far-reaching implications for ML-based error correction.
Finally, we propose a time-efficient and size-invariant pooling for faulty measurement scenarios. Since the lack of effective and efficient error correction is a well-known limiting factor for the development of quantum computers, our contribution can play a role in using machine learning tools for overcoming the current technological limitations of many-qubit systems.

\newpage
\bibliography{aaai24}

\begin{thebibliography}{72}
\providecommand{\natexlab}[1]{#1}

\bibitem[{Aharonov and Ben-Or(1997)}]{aharonov1997fault}
Aharonov, D.; and Ben-Or, M. 1997.
\newblock Fault-tolerant quantum computation with constant error.
\newblock In \emph{Proceedings of the twenty-ninth annual ACM symposium on Theory of computing}, 176--188.

\bibitem[{Andreasson et~al.(2019)Andreasson, Johansson, Liljestrand, and Granath}]{andreasson2019quantum}
Andreasson, P.; Johansson, J.; Liljestrand, S.; and Granath, M. 2019.
\newblock Quantum error correction for the toric code using deep reinforcement learning.
\newblock \emph{Quantum}, 3: 183.

\bibitem[{Ballance et~al.(2016)Ballance, Harty, Linke, Sepiol, and Lucas}]{ballance2016high}
Ballance, C.; Harty, T.; Linke, N.; Sepiol, M.; and Lucas, D. 2016.
\newblock High-fidelity quantum logic gates using trapped-ion hyperfine qubits.
\newblock \emph{Physical review letters}, 117(6): 060504.

\bibitem[{Bengio, L{\'e}onard, and Courville(2013)}]{bengio2013estimating}
Bengio, Y.; L{\'e}onard, N.; and Courville, A. 2013.
\newblock Estimating or propagating gradients through stochastic neurons for conditional computation.
\newblock \emph{arXiv preprint arXiv:1308.3432}.

\bibitem[{Bennatan, Choukroun, and Kisilev(2018)}]{bennatan2018deep}
Bennatan, A.; Choukroun, Y.; and Kisilev, P. 2018.
\newblock Deep learning for decoding of linear codes-a syndrome-based approach.
\newblock In \emph{2018 IEEE International Symposium on Information Theory (ISIT)}, 1595--1599. IEEE.

\bibitem[{Bombin et~al.(2012)Bombin, Andrist, Ohzeki, Katzgraber, and Martin-Delgado}]{bombin2012strong}
Bombin, H.; Andrist, R.~S.; Ohzeki, M.; Katzgraber, H.~G.; and Martin-Delgado, M.~A. 2012.
\newblock Strong resilience of topological codes to depolarization.
\newblock \emph{Physical Review X}, 2(2): 021004.

\bibitem[{Bose and Dowling(1969)}]{bose1969combinatorial}
Bose, R.~C.; and Dowling, T. 1969.
\newblock \emph{Combinatorial mathematics and its applications: proceedings of the conference held at the University of North Carolina at Chapel Hill, April 10-14, 1967}.
\newblock 4. University of North Carolina Press.

\bibitem[{Bravyi et~al.(2018)Bravyi, Englbrecht, K{\"o}nig, and Peard}]{bravyi2018correcting}
Bravyi, S.; Englbrecht, M.; K{\"o}nig, R.; and Peard, N. 2018.
\newblock Correcting coherent errors with surface codes.
\newblock \emph{npj Quantum Information}, 4(1): 1--6.

\bibitem[{Bravyi and Kitaev(1998)}]{bravyi1998quantum}
Bravyi, S.~B.; and Kitaev, A.~Y. 1998.
\newblock Quantum codes on a lattice with boundary.
\newblock \emph{arXiv preprint quant-ph/9811052}.

\bibitem[{Brown et~al.(2020)Brown, Mann, Ryder, Subbiah, Kaplan, Dhariwal, Neelakantan, Shyam, Sastry, Askell et~al.}]{brown2020language}
Brown, T.; Mann, B.; Ryder, N.; Subbiah, M.; Kaplan, J.~D.; Dhariwal, P.; Neelakantan, A.; Shyam, P.; Sastry, G.; Askell, A.; et~al. 2020.
\newblock Language models are few-shot learners.
\newblock \emph{Advances in neural information processing systems}, 33: 1877--1901.

\bibitem[{Brun(2020)}]{QuantumErrorCorrection}
Brun, T.~A. 2020.
\newblock Quantum Error Correction.

\bibitem[{Buchberger et~al.(2020)Buchberger, H{\"a}ger, Pfister, Schmalen et~al.}]{buchberger2020learned}
Buchberger, A.; H{\"a}ger, C.; Pfister, H.~D.; Schmalen, L.; et~al. 2020.
\newblock Learned Decimation for Neural Belief Propagation Decoders.
\newblock \emph{arXiv preprint arXiv:2011.02161}.

\bibitem[{Calderbank and Shor(1996)}]{calderbank1996good}
Calderbank, A.~R.; and Shor, P.~W. 1996.
\newblock Good quantum error-correcting codes exist.
\newblock \emph{Physical Review A}, 54(2): 1098.

\bibitem[{Cammerer et~al.(2017)Cammerer, Gruber, Hoydis, and ten Brink}]{cammerer2017scaling}
Cammerer, S.; Gruber, T.; Hoydis, J.; and ten Brink, S. 2017.
\newblock Scaling deep learning-based decoding of polar codes via partitioning.
\newblock In \emph{GLOBECOM 2017-2017 IEEE Global Communications Conference}, 1--6. IEEE.

\bibitem[{Chamberland and Ronagh(2018)}]{chamberland2018deep}
Chamberland, C.; and Ronagh, P. 2018.
\newblock Deep neural decoders for near term fault-tolerant experiments.
\newblock \emph{Quantum Science and Technology}, 3(4): 044002.

\bibitem[{Choukroun and Wolf(2022)}]{choukroun2022error}
Choukroun, Y.; and Wolf, L. 2022.
\newblock Error Correction Code Transformer.
\newblock \emph{Advances in Neural Information Processing Systems (NeurIPS)}.

\bibitem[{Choukroun and Wolf(2023)}]{choukroun2022denoising}
Choukroun, Y.; and Wolf, L. 2023.
\newblock Denoising Diffusion Error Correction Codes.
\newblock \emph{International Conference on Learning Representations (ICLR)}.

\bibitem[{Cory et~al.(1998)Cory, Price, Maas, Knill, Laflamme, Zurek, Havel, and Somaroo}]{cory1998experimental}
Cory, D.~G.; Price, M.; Maas, W.; Knill, E.; Laflamme, R.; Zurek, W.~H.; Havel, T.~F.; and Somaroo, S.~S. 1998.
\newblock Experimental quantum error correction.
\newblock \emph{Physical Review Letters}, 81(10): 2152.

\bibitem[{Delfosse and Nickerson(2021)}]{delfosse2021almost}
Delfosse, N.; and Nickerson, N.~H. 2021.
\newblock Almost-linear time decoding algorithm for topological codes.
\newblock \emph{Quantum}, 5: 595.

\bibitem[{Dennis et~al.(2002)Dennis, Kitaev, Landahl, and Preskill}]{dennis2002topological}
Dennis, E.; Kitaev, A.; Landahl, A.; and Preskill, J. 2002.
\newblock Topological quantum memory.
\newblock \emph{Journal of Mathematical Physics}, 43(9): 4452--4505.

\bibitem[{Duclos-Cianci and Poulin(2010)}]{duclos2010fast}
Duclos-Cianci, G.; and Poulin, D. 2010.
\newblock Fast decoders for topological quantum codes.
\newblock \emph{Physical review letters}, 104(5): 050504.

\bibitem[{Duclos-Cianci and Poulin(2013)}]{duclos2013fault}
Duclos-Cianci, G.; and Poulin, D. 2013.
\newblock Fault-tolerant renormalization group decoder for abelian topological codes.
\newblock \emph{arXiv preprint arXiv:1304.6100}.

\bibitem[{Edmonds(1965)}]{edmonds1965paths}
Edmonds, J. 1965.
\newblock Paths, trees, and flowers.
\newblock \emph{Canadian Journal of mathematics}, 17: 449--467.

\bibitem[{Fowler(2013)}]{fowler2013minimum}
Fowler, A.~G. 2013.
\newblock Minimum weight perfect matching of fault-tolerant topological quantum error correction in average $ O (1) $ parallel time.
\newblock \emph{arXiv preprint arXiv:1307.1740}.

\bibitem[{Fowler, Whiteside, and Hollenberg(2012)}]{fowler2012towards}
Fowler, A.~G.; Whiteside, A.~C.; and Hollenberg, L.~C. 2012.
\newblock Towards practical classical processing for the surface code.
\newblock \emph{Physical review letters}, 108(18): 180501.

\bibitem[{Foxen et~al.(2020)Foxen, Neill, Dunsworth, Roushan, Chiaro, Megrant, Kelly, Chen, Satzinger, Barends et~al.}]{foxen2020demonstrating}
Foxen, B.; Neill, C.; Dunsworth, A.; Roushan, P.; Chiaro, B.; Megrant, A.; Kelly, J.; Chen, Z.; Satzinger, K.; Barends, R.; et~al. 2020.
\newblock Demonstrating a continuous set of two-qubit gates for near-term quantum algorithms.
\newblock \emph{Physical Review Letters}, 125(12): 120504.

\bibitem[{Gidney(2021)}]{gidney2021stim}
Gidney, C. 2021.
\newblock Stim: a fast stabilizer circuit simulator.
\newblock \emph{{Quantum}}, 5: 497.

\bibitem[{Gottesman(1997)}]{gottesman1997stabilizer}
Gottesman, D. 1997.
\newblock \emph{Stabilizer codes and quantum error correction}.
\newblock California Institute of Technology.

\bibitem[{Greenberger, Horne, and Zeilinger(1989)}]{greenberger1989going}
Greenberger, D.~M.; Horne, M.~A.; and Zeilinger, A. 1989.
\newblock Going beyond Bell’s theorem.
\newblock In \emph{Bell’s theorem, quantum theory and conceptions of the universe}, 69--72. Springer.

\bibitem[{Gruber et~al.(2017)Gruber, Cammerer, Hoydis, and ten Brink}]{gruber2017deep}
Gruber, T.; Cammerer, S.; Hoydis, J.; and ten Brink, S. 2017.
\newblock On deep learning-based channel decoding.
\newblock In \emph{2017 51st Annual Conference on Information Sciences and Systems (CISS)}, 1--6. IEEE.

\bibitem[{Higgott(2022)}]{higgott2022pymatching}
Higgott, O. 2022.
\newblock PyMatching: A Python package for decoding quantum codes with minimum-weight perfect matching.
\newblock \emph{ACM Transactions on Quantum Computing}, 3(3): 1--16.

\bibitem[{Higgott and Gidney(2022)}]{pymatchingv2}
Higgott, O.; and Gidney, C. 2022.
\newblock PyMatching v2.
\newblock \url{https://github.com/oscarhiggott/PyMatching}.

\bibitem[{Huang, Newman, and Brown(2020)}]{huang2020fault}
Huang, S.; Newman, M.; and Brown, K.~R. 2020.
\newblock Fault-tolerant weighted union-find decoding on the toric code.
\newblock \emph{Physical Review A}, 102(1): 012419.

\bibitem[{Huang et~al.(2019)Huang, Yang, Chan, Tanttu, Hensen, Leon, Fogarty, Hwang, Hudson, Itoh et~al.}]{huang2019fidelity}
Huang, W.; Yang, C.; Chan, K.; Tanttu, T.; Hensen, B.; Leon, R.; Fogarty, M.; Hwang, J.; Hudson, F.; Itoh, K.~M.; et~al. 2019.
\newblock Fidelity benchmarks for two-qubit gates in silicon.
\newblock \emph{Nature}, 569(7757): 532--536.

\bibitem[{Hutter, Wootton, and Loss(2014)}]{hutter2014efficient}
Hutter, A.; Wootton, J.~R.; and Loss, D. 2014.
\newblock Efficient Markov chain Monte Carlo algorithm for the surface code.
\newblock \emph{Physical Review A}, 89(2): 022326.

\bibitem[{Kim et~al.(2018)Kim, Jiang, Rana, Kannan, Oh, and Viswanath}]{kim2018communication}
Kim, H.; Jiang, Y.; Rana, R.; Kannan, S.; Oh, S.; and Viswanath, P. 2018.
\newblock Communication algorithms via deep learning.
\newblock In \emph{Sixth International Conference on Learning Representations (ICLR)}.

\bibitem[{Kingma and Ba(2014)}]{kingma2014adam}
Kingma, D.~P.; and Ba, J. 2014.
\newblock Adam: A method for stochastic optimization.
\newblock \emph{arXiv preprint arXiv:1412.6980}.

\bibitem[{Kitaev(1997{\natexlab{a}})}]{kitaev1997quantum}
Kitaev, A.~Y. 1997{\natexlab{a}}.
\newblock Quantum computations: algorithms and error correction.
\newblock \emph{Russian Mathematical Surveys}, 52(6): 1191.

\bibitem[{Kitaev(1997{\natexlab{b}})}]{kitaev1997quantum3}
Kitaev, A.~Y. 1997{\natexlab{b}}.
\newblock Quantum computations: algorithms and error correction.
\newblock \emph{Russian Mathematical Surveys}, 52(6): 1191.

\bibitem[{Kitaev(1997{\natexlab{c}})}]{kitaev1997quantum2}
Kitaev, A.~Y. 1997{\natexlab{c}}.
\newblock Quantum error correction with imperfect gates.
\newblock In \emph{Quantum communication, computing, and measurement}, 181--188. Springer.

\bibitem[{Kitaev(2003)}]{kitaev2003fault}
Kitaev, A.~Y. 2003.
\newblock Fault-tolerant quantum computation by anyons.
\newblock \emph{Annals of Physics}, 303(1): 2--30.

\bibitem[{Kolmogorov(2009)}]{kolmogorov2009blossom}
Kolmogorov, V. 2009.
\newblock Blossom V: a new implementation of a minimum cost perfect matching algorithm.
\newblock \emph{Mathematical Programming Computation}, 1(1): 43--67.

\bibitem[{Krastanov and Jiang(2017)}]{krastanov2017deep}
Krastanov, S.; and Jiang, L. 2017.
\newblock Deep neural network probabilistic decoder for stabilizer codes.
\newblock \emph{Scientific reports}, 7(1): 1--7.

\bibitem[{Kuo and Lu(2020)}]{kuo2020hardnesses}
Kuo, K.-Y.; and Lu, C.-C. 2020.
\newblock On the hardnesses of several quantum decoding problems.
\newblock \emph{Quantum Information Processing}, 19(4): 1--17.

\bibitem[{Lidar and Brun(2013)}]{lidar2013quantum}
Lidar, D.~A.; and Brun, T.~A. 2013.
\newblock \emph{Quantum error correction}.
\newblock Cambridge university press.

\bibitem[{Lin et~al.(2021)Lin, Wang, Liu, and Qiu}]{lin2021survey}
Lin, T.; Wang, Y.; Liu, X.; and Qiu, X. 2021.
\newblock A survey of transformers.
\newblock \emph{arXiv preprint arXiv:2106.04554}.

\bibitem[{Lugosch and Gross(2017)}]{lugosch2017neural}
Lugosch, L.; and Gross, W.~J. 2017.
\newblock Neural offset min-sum decoding.
\newblock In \emph{2017 IEEE International Symposium on Information Theory (ISIT)}, 1361--1365. IEEE.

\bibitem[{MacKay(2003)}]{mackay2003information}
MacKay, D.~J. 2003.
\newblock \emph{Information theory, inference and learning algorithms}.
\newblock Cambridge university press.

\bibitem[{Meinerz, Park, and Trebst(2022)}]{meinerz2022scalable}
Meinerz, K.; Park, C.-Y.; and Trebst, S. 2022.
\newblock Scalable neural decoder for topological surface codes.
\newblock \emph{Physical Review Letters}, 128(8): 080505.

\bibitem[{Micali and Vazirani(1980)}]{micali1980v}
Micali, S.; and Vazirani, V.~V. 1980.
\newblock An O (v| v| c| E|) algoithm for finding maximum matching in general graphs.
\newblock In \emph{21st Annual Symposium on Foundations of Computer Science (sfcs 1980)}, 17--27. IEEE.

\bibitem[{Nachmani, Be'ery, and Burshtein(2016)}]{nachmani2016learning}
Nachmani, E.; Be'ery, Y.; and Burshtein, D. 2016.
\newblock Learning to decode linear codes using deep learning.
\newblock In \emph{2016 54th Annual Allerton Conference on Communication, Control, and Computing (Allerton)}, 341--346. IEEE.

\bibitem[{Nachmani and Wolf(2019)}]{nachmani2019hyper}
Nachmani, E.; and Wolf, L. 2019.
\newblock Hyper-graph-network decoders for block codes.
\newblock In \emph{Advances in Neural Information Processing Systems}, 2326--2336.

\bibitem[{Neumann, Wigner, and Hofstadter(1955)}]{neumann1955mathematical}
Neumann, J.; Wigner, E.~P.; and Hofstadter, R. 1955.
\newblock \emph{Mathematical foundations of quantum mechanics}.
\newblock Princeton university press.

\bibitem[{Nielsen and Chuang(2002)}]{nielsen2002quantum}
Nielsen, M.~A.; and Chuang, I. 2002.
\newblock Quantum computation and quantum information.

\bibitem[{Park and Meinerz(2022)}]{park2022scalablecode}
Park, C.-Y.; and Meinerz, K. 2022.
\newblock Open-source C++ implementation of the Union-Find decoder, \url{https://github.com/chaeyeunpark/UnionFind}.
\newblock \emph{Physical Review Letters}, 128(8): 080505.

\bibitem[{Pearl(1988)}]{pearl1988probabilistic}
Pearl, J. 1988.
\newblock \emph{Probabilistic reasoning in intelligent systems: networks of plausible inference}.
\newblock Morgan kaufmann.

\bibitem[{Preskill(1998)}]{preskill1998reliable}
Preskill, J. 1998.
\newblock Reliable quantum computers.
\newblock \emph{Proceedings of the Royal Society of London. Series A: Mathematical, Physical and Engineering Sciences}, 454(1969): 385--410.

\bibitem[{Raimond and Haroche(1996)}]{raimond1996quantum}
Raimond, J.; and Haroche, S. 1996.
\newblock Quantum computing: dream or nightmare.
\newblock \emph{DARK MATTER IN COSMOLOGY QUANTUM MEASUREMENTS EXPERIMENTAL GRA VITA Tl ON}, 341.

\bibitem[{Schindler et~al.(2011)Schindler, Barreiro, Monz, Nebendahl, Nigg, Chwalla, Hennrich, and Blatt}]{schindler2011experimental}
Schindler, P.; Barreiro, J.~T.; Monz, T.; Nebendahl, V.; Nigg, D.; Chwalla, M.; Hennrich, M.; and Blatt, R. 2011.
\newblock Experimental repetitive quantum error correction.
\newblock \emph{Science}, 332(6033): 1059--1061.

\bibitem[{Shannon(1948)}]{shannon1948mathematical}
Shannon, C.~E. 1948.
\newblock A mathematical theory of communication.
\newblock \emph{The Bell system technical journal}, 27(3): 379--423.

\bibitem[{Shor(1995)}]{shor1995scheme}
Shor, P.~W. 1995.
\newblock Scheme for reducing decoherence in quantum computer memory.
\newblock \emph{Physical review A}, 52(4): R2493.

\bibitem[{Sweke et~al.(2020)Sweke, Kesselring, van Nieuwenburg, and Eisert}]{sweke2020reinforcement}
Sweke, R.; Kesselring, M.~S.; van Nieuwenburg, E.~P.; and Eisert, J. 2020.
\newblock Reinforcement learning decoders for fault-tolerant quantum computation.
\newblock \emph{Machine Learning: Science and Technology}, 2(2): 025005.

\bibitem[{Torlai and Melko(2017)}]{torlai2017neural}
Torlai, G.; and Melko, R.~G. 2017.
\newblock Neural decoder for topological codes.
\newblock \emph{Physical review letters}, 119(3): 030501.

\bibitem[{Varona and Martin-Delgado(2020)}]{varona2020determination}
Varona, S.; and Martin-Delgado, M.~A. 2020.
\newblock Determination of the semion code threshold using neural decoders.
\newblock \emph{Physical Review A}, 102(3): 032411.

\bibitem[{Varsamopoulos, Criger, and Bertels(2017)}]{varsamopoulos2017decoding}
Varsamopoulos, S.; Criger, B.; and Bertels, K. 2017.
\newblock Decoding small surface codes with feedforward neural networks.
\newblock \emph{Quantum Science and Technology}, 3(1): 015004.

\bibitem[{Vaswani et~al.(2017)Vaswani, Shazeer, Parmar, Uszkoreit, Jones, Gomez, Kaiser, and Polosukhin}]{vaswani2017attention}
Vaswani, A.; Shazeer, N.; Parmar, N.; Uszkoreit, J.; Jones, L.; Gomez, A.~N.; Kaiser, {\L}.; and Polosukhin, I. 2017.
\newblock Attention is all you need.
\newblock In \emph{Advances in neural information processing systems}, 5998--6008.

\bibitem[{Wagner, Kampermann, and Bru{\ss}(2020)}]{wagner2020symmetries}
Wagner, T.; Kampermann, H.; and Bru{\ss}, D. 2020.
\newblock Symmetries for a high-level neural decoder on the toric code.
\newblock \emph{Physical Review A}, 102(4): 042411.

\bibitem[{Wang, Harrington, and Preskill(2003)}]{wang2003confinement}
Wang, C.; Harrington, J.; and Preskill, J. 2003.
\newblock Confinement-Higgs transition in a disordered gauge theory and the accuracy threshold for quantum memory.
\newblock \emph{Annals of Physics}, 303(1): 31--58.

\bibitem[{Wang et~al.(2020)Wang, Li, Khabsa, Fang, and Ma}]{wang2020linformer}
Wang, S.; Li, B.~Z.; Khabsa, M.; Fang, H.; and Ma, H. 2020.
\newblock Linformer: Self-attention with linear complexity.
\newblock \emph{arXiv preprint arXiv:2006.04768}.

\bibitem[{Wootters and Zurek(1982)}]{wootters1982single}
Wootters, W.~K.; and Zurek, W.~H. 1982.
\newblock A single quantum cannot be cloned.
\newblock \emph{Nature}, 299(5886): 802--803.

\bibitem[{Wootton and Loss(2012)}]{wootton2012high}
Wootton, J.~R.; and Loss, D. 2012.
\newblock High threshold error correction for the surface code.
\newblock \emph{Physical review letters}, 109(16): 160503.

\bibitem[{Xiong et~al.(2020)}]{xiong2020layer}
Xiong, R.; et~al. 2020.
\newblock On layer normalization in the transformer architecture.
\newblock \emph{arXiv:2002.04745}.

\end{thebibliography}

\appendix

\section{A. Classical vs. Quantum Error Correction}

\begin{figure}[h]
\centering
\begin{tabular}{c}
\includegraphics[trim={0 0 0 0},clip,width=1\linewidth]{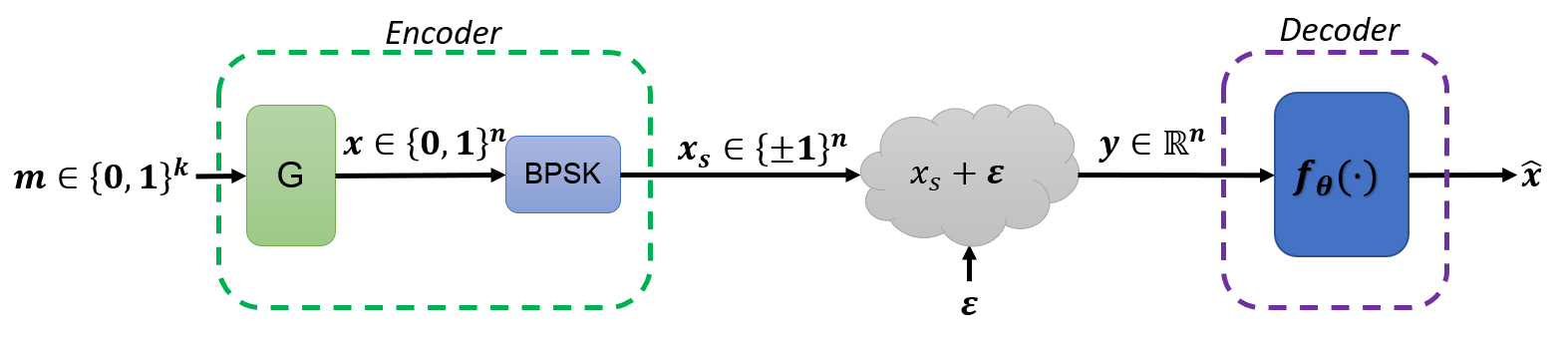}\\
(a) \\
\includegraphics[trim={0 0 0 0},clip,width=1\linewidth]{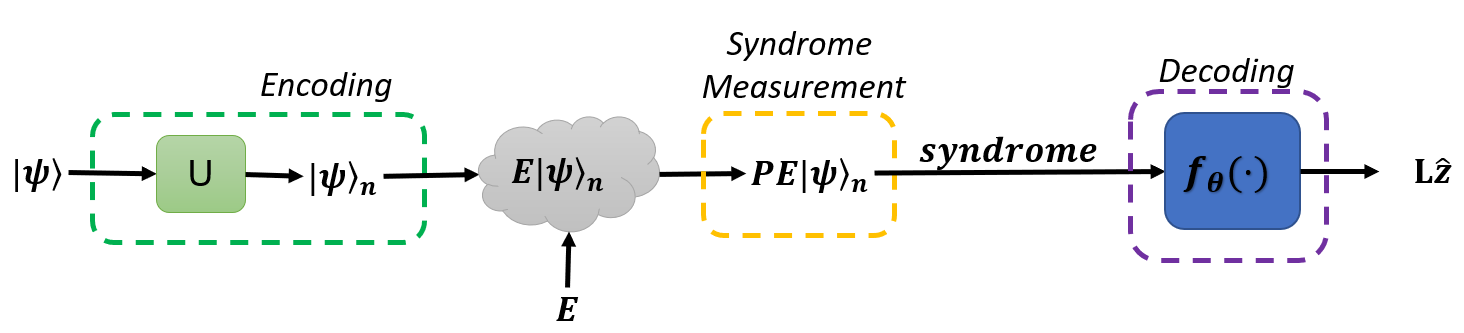}\\
(b) 
\end{tabular}
\caption{A detailed illustration of the (a) classical and (b) quantum ECC system. Our work focuses on the design and training of the parameterized quantum decoder.}
\label{fig:qecc_illustration}
\end{figure}

We present in Fig.~\ref{fig:qecc_illustration} a more detailed version of Fig.1 in the main paper, illustrating the differences between classical and quantum error correction. These differences are detailed in the main text.

\section{B. MLP Decoder}
\label{sec:apendix_MLP}
We present below the results obtained with Fully-Connected (FC) models on the L=6 and L=8 codes under the same training and capacity budgets as the QECCT. 
We present the best FC results, obtained using 10 layers and d=256-dimensional embedding on high SNRs.

These results are obtained after experimenting with different architecture parameters. 
The use of CNN and RNN is not well defined in our one-dimensional input setting and thus not experimented with.

\begin{figure}[h]
\centering
\noindent  \begin{tabular}{@{}ccc@{}}
  \includegraphics[trim={0 0 0 0},clip, width=0.49\linewidth]{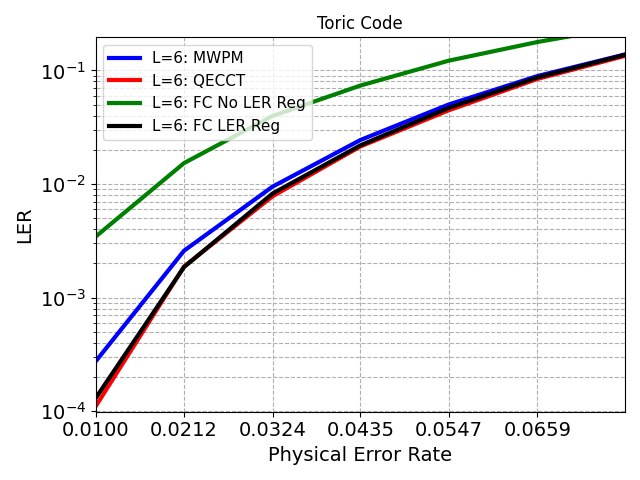}&
    \includegraphics[trim={0 0 0 0},clip, width=0.49\linewidth]{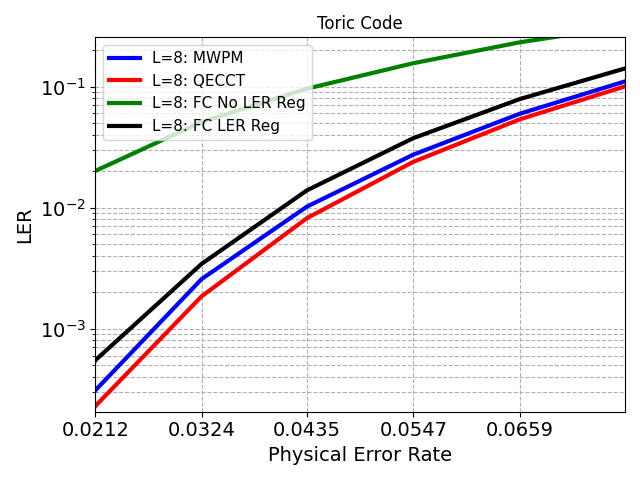}&
      \end{tabular}
\vspace{-1mm}
  \caption{LER and BER performance for various physical error rates and lattice length on the Toric code with \emph{independent} noise and \emph{without} faulty syndrome measurements.}
\label{fig:appendix_Toric_norep_indepedent}
\end{figure}

\begin{figure}[t]
\centering
\noindent  \begin{tabular}{@{}ccc@{}}
  \includegraphics[trim={0 0 0 0},clip, width=0.49\linewidth]{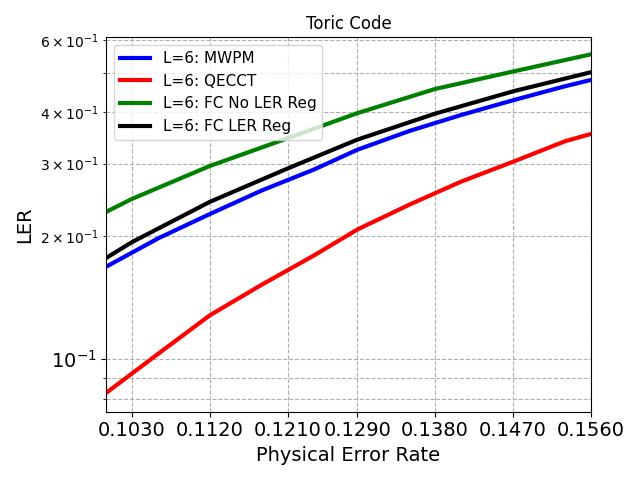}&
    \includegraphics[trim={0 0 0 0},clip, width=0.49\linewidth]{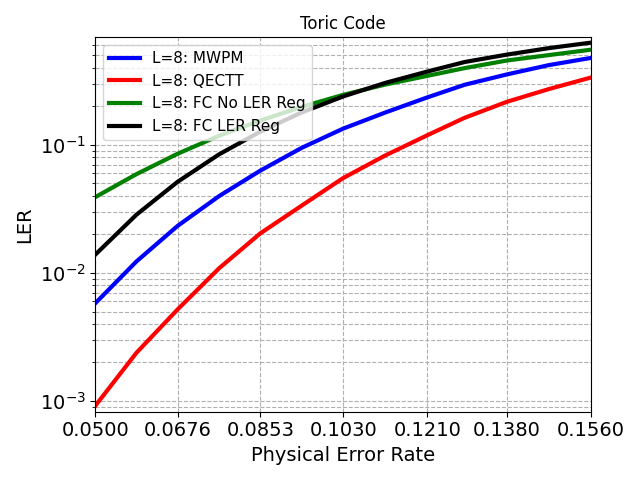}&
      \end{tabular}
\vspace{-1mm}
  \caption{LER and BER performance for various physical error rates and lattice length on the Toric code with \emph{depolarization} noise and \emph{without} faulty syndrome measurements.}
\label{fig:appendix_Toric_norep_depolar}
\end{figure}

We can observe in Figures \ref{fig:appendix_Toric_norep_indepedent} and \ref{fig:appendix_Toric_norep_depolar} that our model outperforms the deep FCNN and is especially suitable for larger codes and depolarization noise where the correlation between the X and Z errors can be considered via the pairwise self-attention mechanism.
Also, we can observe the advantage of the LER regularization which can bring orders of magnitude in the improvement of the accuracy and make the MLP model comparable to MWPM.

\section{C. Impact of the noise estimator $g_{\omega}$}
\begin{figure}[t]
\centering
\noindent  \begin{tabular}{@{}ccc@{}}
  \includegraphics[trim={0 0 0 0},clip, width=0.49\linewidth]{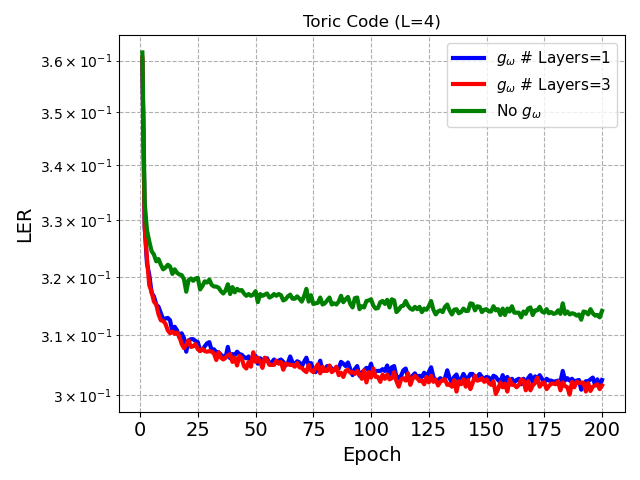}&
    \includegraphics[trim={0 0 0 0},clip, width=0.49\linewidth]{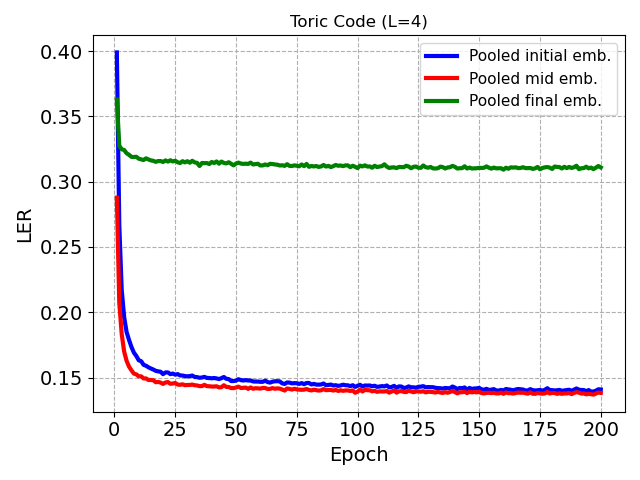}&
    \\
    (a) & (b) &
      \end{tabular}
  \caption{Impact of the $g_{\omega}$ on LER performance during training with an $N=6,d=128$ architecture (a). 
  "No $g_{\omega}$" refers to constant mapping, i.e., $g_{\omega}={1}_{n}$. 
  Impact of pooling (averaging) performed at different levels of the network for a $N=6,d=128$ model (b). The average final testing LER of mid-level pooling is 5\% lower than the initial embedding pooling.}
\label{fig:ablation_g}
\end{figure}

Figure \ref{fig:ablation_g}(a) explores the impact of $g_{\omega}$ on performance. Since in order to use the QECCT the initial scaling estimate needs to be defined, we compare our approach to constant scaling with ones, meaning $g_{\omega}:={1}_{n}$. We further explore two architectures for  $g_{\omega}$: one with a single hidden layer and one with three hidden layers. 
Evidently, the initial noise estimator is critical for performance. However, the network architecture is less impactful. Given more resources, one can check the impact of additional architectures. 



\section{D. Impact of the various objectives}
\begin{figure}[t]
\centering
\noindent  \begin{tabular}{@{}ccc@{}}
  \includegraphics[trim={0 0 0 0},clip, width=0.49\linewidth]{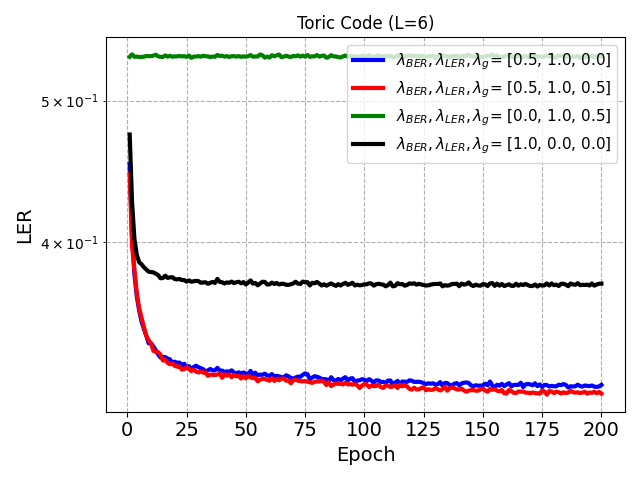}&
    \includegraphics[trim={0 0 0 0},clip, width=0.49\linewidth]{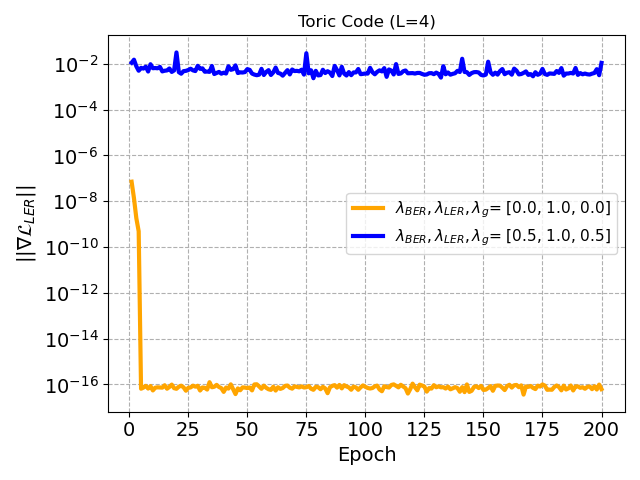}&
    \\
    (a) & (b) &
      \end{tabular}
  \caption{Impact of the different loss terms on performance (a). Impact of regularization on the training dynamics (b) with a $N=6,d=128$ model.}
\label{fig:ablation_LER}
\end{figure}

Figure \ref{fig:ablation_LER}(a) depicts the impact of the different loss terms of the training objective. 
First, we can observe that training solely with the $\mathcal{L}_{LER}$ objective converges rapidly to a bad local minimum. This is illustrated in Figure \ref{fig:ablation_LER}(b) where we display the gradient norm average over the network layers.

We hypothesize that this issue may be due to the fact that the dimension of the LER objective is rather small and the gradients can collapse in the product of the code elements in Eq 9, since $\sigma(x)^{n}\xrightarrow[n \to \infty]{} 0$. 
{\color{black}
Also, the product of the elements allows the creation of a saddle point objective.
}

Evidently, training with $\mathcal{L}_{BER}$ only produces far worse results than combining it with the $\mathcal{L}_{LER}$ objective.
For high SNR, a model trained with $\mathcal{L}_{BER}$ objective yields a {\color{black}27 times higher LER} than the model trained with combined objectives.
Moreover, optimizing with the noise estimator objective $\mathcal{L}_{g}$ results, for high SNR, in a 46\% improvement over not employing regularization. 

\section{E. Impact of Pooling}
Figure \ref{fig:ablation_g}(b) depicts the impact of different pooling (averaging) scenarios. Pooling is performed either after the initial embedding, in the middle of the network, or over the final embedding. 
Evidently, performing pooling in the middle layer allows better decoding. 
Pooling the final embedding (similarly to voting) is not effective.
\begin{figure}[t]
\centering
\noindent  \begin{tabular}{@{}ccc@{}}
  \includegraphics[trim={0 0 0 0},clip, width=0.49\linewidth]{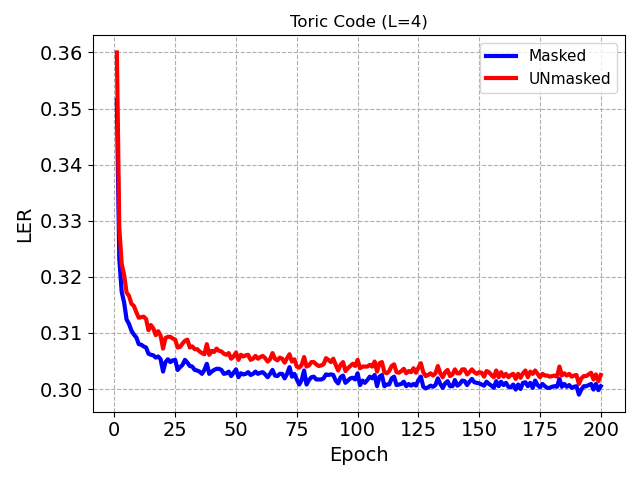}&
    \includegraphics[trim={0 0 0 0},clip, width=0.49\linewidth]{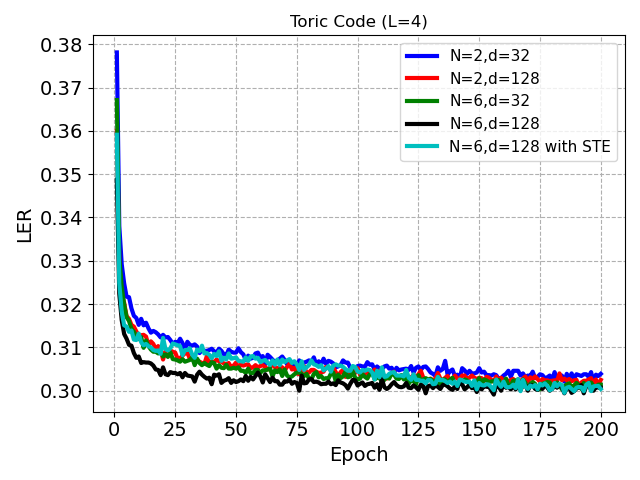}&\\
    (a) & (b) &
      \end{tabular}
  \caption{(a) Impact of masking on accuracy with an \mbox{$N=2,d=128$} model. (b) Impact of the model size on accuracy with an $N=6,d=128$ model. We also provide the performance of the STE as the ${bin}$ function.}
\label{fig:ablation_mask_arch}
\end{figure}

\section{F. Impact of the Mask and the Architecture}
Finally, we present the impact of masking in Figure \ref{fig:ablation_mask_arch}(a) and the impact of the model's capacity and the performance of the STE as ${bin}$ function in Figure \ref{fig:ablation_mask_arch}(b).
We can observe that while less important than with classical codes \citep{choukroun2022error}, the mask still substantially impacts performance.
Also, we note that increasing the capacity of the network enables better representation and decoding. 

\section{G. Toric Codes}
\begin{figure}[t]
\centering
  \includegraphics[trim={0 0 0 0},clip, width=0.79\linewidth]{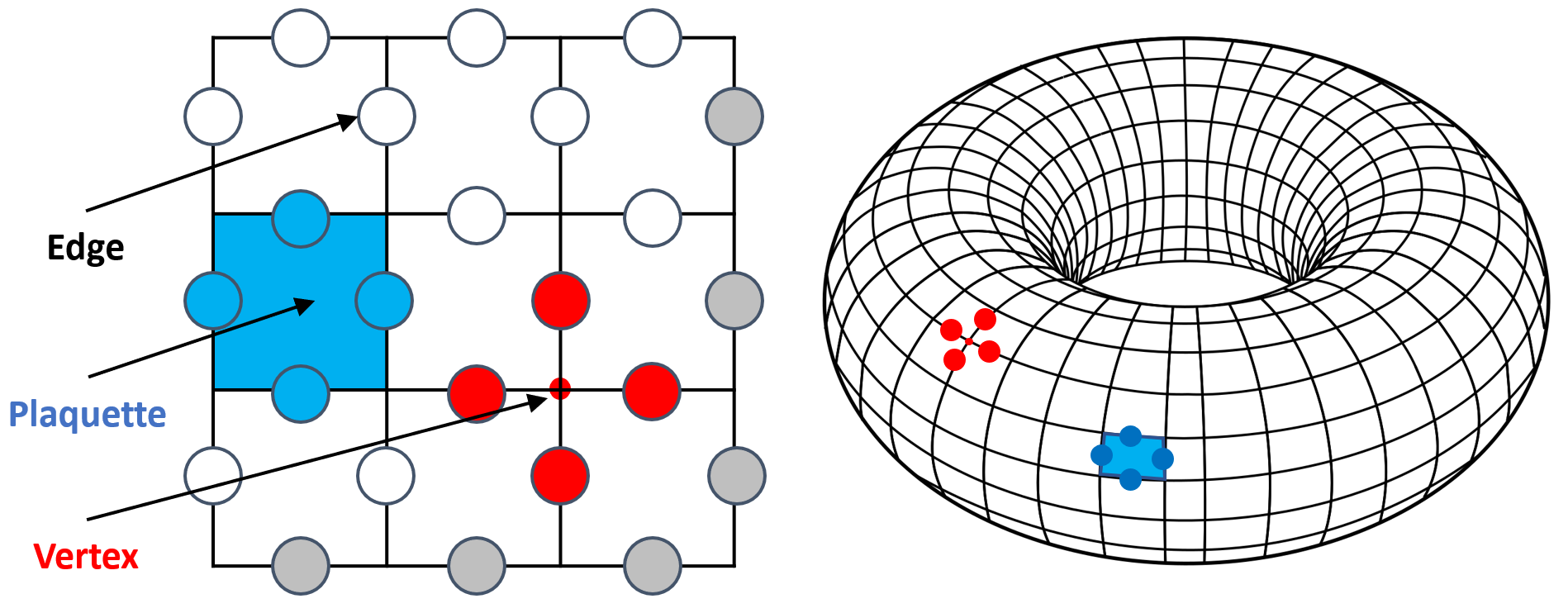}
  \caption{The lattice representation of the (k=3) Toric code and its embedding into a three-dimensional torus.
  Gray edges represent periodic boundaries.
}
\label{fig:illus-toric}
\end{figure}

We provide an illustration of the Toric code in Figure \ref{fig:illus-toric}.

\end{document}